\documentclass[12pt]{iopart}
\usepackage{epsfig}
\usepackage{graphicx}

\newcommand{\be}{\begin{equation}}
\newcommand{\ee}{\end{equation}}
\newcommand{\bd}{\begin{displaymath}}
\newcommand{\ed}{\end{displaymath}}

\newcommand{\bra}{\langle}
\newcommand{\ket}{\rangle}
\newcommand{\bigbra}{\left\langle}
\newcommand{\bigket}{\right\rangle}
\newcommand{\order}{{\cal O}}

\newcommand{\minus}{{\!-\!}}
\newcommand{\sgn}{{\rm sgn}}

\newcommand{\R}{{\rm I\!R}}

\newcommand{\bsigma}{{\mbox{\boldmath $\sigma$}}}

\newcommand{\bc}{\ensuremath{\mathbf{c}}}
\newcommand{\bh}{\ensuremath{\mathbf{h}}}

\newcommand{\bm}{\ensuremath{\mathbf{m}}}
\newcommand{\bq}{\ensuremath{\mathbf{q}}}

\newcommand{\bx}{\ensuremath{\mathbf{x}}}

\usepackage{epsfig}
\usepackage{graphicx}

\newcommand{\bxi}{{\mbox{\boldmath $\xi$}}}

\newcommand{\betaslow}{\tilde{\beta}}
\newcommand{\btau}{{\mbox{\boldmath $\tau$}}}

\begin{document}

\title[Slowly evolving geometry in recurrent neural networks I]{Slowly evolving geometry in recurrent neural networks I:
the extreme dilution regime}
\author{B Wemmenhove$^\dag$, N S  Skantzos$^\ddag$ and A C C Coolen$^S$}

\address{\dag ~Institute for
Theoretical Physics, University of Amsterdam,\\ ~~
Valckenierstraat 65, 1018 XE Amsterdam, The Netherlands}

\address{\ddag ~ Departament de F\'{\i}sica Fonamental, Facultat de F\'{\i}sica, Universitat
de Barcelona,\\ ~~ 08028 Barcelona, Spain}

\address{\S ~ Department of Mathematics, King's College London,\\ ~~ The Strand,
London WC2R 2LS, United Kingdom}

\begin{abstract}
We study extremely diluted spin models of neural networks in which
the connectivity evolves in time, although adiabatically slowly
compared to the neurons, according to stochastic equations which
on average aim to reduce frustration. The (fast) neurons and
(slow) connectivity variables equilibrate separately, but at
different temperatures. Our model is exactly solvable in
equilibrium. We obtain phase diagrams upon making the condensed
ansatz (i.e. recall of one pattern). These show that, as the
connectivity temperature is lowered, the volume of the retrieval
phase diverges and the fraction of mis-aligned spins is reduced.
Still one always retains a region in the retrieval phase where
recall states other than the one corresponding to the `condensed'
pattern are locally stable, so the associative memory character of
our model is preserved.
\end{abstract}

\pacs{75.10.Nr, 05.20.-y, 64.60.Cn}

\ead{\tt wemmenho@science.uva.nl, nikos@ffn.ub.es,
tcoolen@mth.kcl.ac.uk}


\section{Introduction}

Most statistical mechanical studies of recurrent neural networks
have traditionally been concerned with systems in which the
dynamical variables are either the neurons (see e.g.
\cite{Hopfield,AGS85a,AGS85b,DGZ} or the reviews
\cite{Statics,Dynamics} and references therein), or their
interactions (or synapses, see e.g.
\cite{Gardner,HKT,BiehlSchwarze,KinouchiCaticha} or the reviews
\cite{KinzelOpper,Watkinetal,MaceCoolen} and references therein).
The first type of processes describe network operation, whereas
the second correspond to learning. These areas have by now been
investigated quite extensively. In contrast, only a modest number
of studies involved coupled dynamical laws for both neurons and
interactions
\cite{Shinomoto,CoolenSherrington,PCS1,FeldmanDotsenko,DotsFranzMezard,JBC,JABCP,CoolenUezu},
to reflect the complex dynamical interplay between synapses and
neurons found in the real brain. The approach usually adopted in
the latter studies, to obtain analytically solvable models, is the
introduction of a hierarchy of adiabatically separated
time-scales, such that the fast variables (taken to be the
neurons) are in equilibrium on the time-scales where the slow
variables (the interactions, taken to be symmetric) evolve. One
can also introduce further levels in the hierarchy by introducing
different classes of interactions, each evolving on different
characteristic time-scales \cite{CoolenUezu}. The resulting
formalism  involves
 nested replica theories, with Parisi matrices \cite{MPV}
 in which the number of blocks at
each level is the ratio of temperatures of subsequent levels in
the hierarchy of equilibrating degrees of freedom. Such models can
also served to derive Parisi's replica symmetry breaking scheme
\cite{MourikCoolen}. In neural network studies the dynamics of the
interactions have usually been governed by Langevin equations in
which the deterministic forces are proportional to expectation
values of neuronal pair correlations (with the neuron state
statistics corresponding to Boltzmann equilibrium, given the
instantaneous values of the interactions), potentially biased to
reflect the possibility of recall of a pattern. In
\cite{FeldmanDotsenko,DotsFranzMezard} the interactions were taken
to  evolve away from an initial state given by Hopfield's
\cite{Hopfield} interaction matrix, with an extensive number of
stored patterns. There it was found that for low interaction
temperatures  the network collapsed into an undesirable  so-called
`super-ferromagnetic' state, whereas for negative replica
dimension (corresponding to anti-Hebbian learning)  the storage
capacity of the network was found to be enhanced.

All papers dealing with the theory of coupled neuronal and
interaction dynamics  published so far assumed full connectivity:
each neuron interacted which each other neuron, with the magnitude
and sign of the interactions evolving in time.  Here we propose
and study a model of a symmetrically diluted recurrent neural
network in which the {\em geometry} (or connectivity) is allowed
to change slowly.
 On time-scales where  the
neuron variables are in thermodynamic equilibrium, the microscopic
realization of the (discrete) dilution variables (reflecting the
connectivity) is allowed to evolve slowly and stochastically,
driven by forces aiming at a reduction of global frustration,
without however changing the actual {\em values} of the bonds (the
latter are frozen, given by Hopfield's \cite{Hopfield} recipe). It
has been known that one may store information in recurrent neural
networks solely by eliminating frustrated bonds, but this has
always been done by hand (see e.g. \cite{JonkerCoolen} and
references therein). Here the system is allowed to adapt its
geometry autonomously. It should be emphasized that there is an
important difference between having dynamic bonds with Hebbian
type dynamical laws, as in
\cite{Shinomoto,CoolenSherrington,PCS1,JBC,JABCP,CoolenUezu}, and
the present situation of having dynamic geometry with fixed
Hebbian values for active bonds. The former definitions imply
irreversible modification or even elimination of stored
information, whereas in the present paper, since the {\em values}
assigned to the active bonds are not modified, the slow adaptation
is fully reversible (one can always return to random dilution) and
all stored information is retained.

The scaling with the system size $N$ chosen for the average
connectivity $c$ in the system (the average number of bonds per
spin) will have a strong influence on the structure of the
resulting theory. In this first paper we consider the so-called
`extreme dilution' regime
 \cite{WS}, defined by $\lim_{N\to\infty}c^{-1}=\lim_{N\to\infty}c/N=0$
 (a second paper will be devoted to the finite connectivity
 regime, where $c=\order(N^0)$ as $N\to\infty$).
We solve our coupled spin and geometry dynamics model
analytically, in replica symmetric ansatz. We find that, as a
result of the connectivity adaptation, the network geometry
becomes more ordered to boost retrieval of condensed patterns, as
a result of which the system's retrieval phase is enhanced
compared to that of the corresponding network with a quenched
random connectivity matrix as studied in \cite{WS}, and that the
fraction of `misaligned spins' is reduced as the temperature of
the connectivity variables is lowered. Yet one still retains
regions in the phase diagram where the alternative (presently
non-condensed) pure retrieval states remain locally stable, so
that the system continues to function as an associative memory.

\section{Definitions}

We study diluted Hopfield \cite{Hopfield} type recurrent neural
networks, with (fast) binary neurons $\sigma_i \in \{-1,1\}$
(denoting quiet versus firing states) and $i=1,\ldots, N$. The
geometry of the system is defined by connectivity variables
$c_{ij} \in \{0,1\}$, with $c_{ji} = c_{ij}$ and $c_{ii} = 0$. Our
neurons evolve according to Glauber-type local field alignment at
temperature $T=\beta^{-1}$, with the fields defined by $h_i=\sum_j
\frac{c_{ij}}{c} \sum_{\mu = 1}^p \xi_i^\mu \xi_j^\mu \sigma_j$,
i.e. with Hebbian interactions whenever $c_{ij}=1$ (when a bond
$(i,j)$ is present). For frozen geometry $\{c_{ij}\}$ our Ising
spin  neurons would equilibrate to a Boltzmann state characterized
by the Hamiltonian
\begin{equation}
H_{\rm f}(\bsigma,\bc) = -\sum_{i<j} \frac{c_{ij}}{c} \sum_{\mu =
1}^p \xi_i^\mu \xi_j^\mu \sigma_i \sigma_j \label{eq:fastham}
\end{equation}
Here $\bsigma=(\sigma_1,\ldots,\sigma_N)$ and $\bc=\{c_{ij}\}$.
 The $\{ \xi_i^\mu \} \in \{-1,1\}$ with
$\mu= 1, \ldots, p$ represent $p$ fixed patterns
$\bxi^\mu=(\xi_1^\mu,\ldots,\xi_N^\mu)$ to be stored and hopefully
recalled. Instead of frozen, we now take our geometry
 to also evolve in time, albeit on
time-scales much larger than those of the neuronal relaxation (so
the neurons can always be assumed in equilibrium, given the
instantaneous geometry). This  slow process is again taken to be
of a Glauber type, but at temperature
$\tilde{T}=\tilde{\beta}^{-1}$ and  with the connectivity
Hamiltonian
\begin{eqnarray}
H_{\rm s}(\bc) &=& -\frac{1}{\beta} \log Z_{\rm f}(\bc) +
\frac{1}{\betaslow} \log\left(\frac{N}{c}\right) \sum_{i<j} c_{ij}
\label{eq:slowham}
\\
Z_{\rm f}(\bc) &=& \sum_{\bsigma}e^{-\beta H_{\rm f}(\bsigma,\bc)}
\end{eqnarray}
The second term in (\ref{eq:slowham}) acts as a chemical
potential, ensuring an average number of $c$ connections per
neuron.  The pre-factor $1/\betaslow$ will be found helpful later.

The properties of our system at the largest time-scales, where
also the geometry has equilibrated, are characterized by the
partition sum of the slow variables:
\begin{equation}
Z_{\rm s} = \sum_{\bc} e^{-\betaslow H_{\rm s}(\bc)} = \sum_{\bc}
\left[Z_{\rm f}(\bc)\right]^{\betaslow/\beta}
e^{-\log(\frac{N}{c}) \sum_{i<j}c_{ij}} \label{eq:partsumslow}
\end{equation}
This sum is interpreted as describing $n=\tilde{\beta}/\beta$
replicated copies of the fast system, leading to a replica theory
with finite replica dimension $n$. Minimization of
 $H_{\rm s}(\bc)$ should give a `smart' arrangement of the
geometry $\{c_{ij}\}$, taylored to the realization of the
patterns, but constrained to give an average connectivity $c$.  In
the remainder of this paper we calculate phase diagrams and the
fraction of mis-aligned spins. Phases are characterized by the
values of the replicated overlap and  spin glass order parameters
\begin{equation}
m_\alpha^\mu = \lim_{N\to\infty}N^{-1}\sum_i\overline{\bra
\xi^\mu_i \sigma^\alpha_i \ket} ~~~~~~~~ q_{\alpha\beta} =
\lim_{N\to\infty}N^{-1}\sum_i\overline{\bra \sigma^\alpha_i
\sigma^{\beta}_i \ket}
\end{equation}
Here $\overline{\bra \ldots \ket}$ denotes averaging over all
geometries $\{c_{ij}\}$ and all spin-configurations
$\bsigma^\alpha$ in each of the replicas $\alpha=1,\ldots,n$, with
the Boltzmann measure associated with  (\ref{eq:partsumslow}):
\begin{eqnarray}
\overline{\bra G(\{ \bsigma^\alpha \}, \bc) \ket} &=& Z_{\rm
s}^{-1} \sum_{\{ \bsigma^\alpha \} } \sum_{\bc} G(\{
\bsigma^\alpha \}, \bc) \nonumber \\ && \times \exp\left\{
\frac{\beta}{c} \sum_{i<j} c_{ij} \sum_\mu \xi_i^\mu \xi_j^\mu
\sum_{\alpha=1}^n \sigma_i^\alpha \sigma_j^\alpha -
\log\left(\frac{N}{c}\right)\sum_{i<j} c_{ij} \right\}
\label{eq:all_averages}
\end{eqnarray}

\section{Equilibrium analysis}

\subsection{Calculation of the RS free energy}

The thermodynamic properties of the stationary state, with
equilibrated geometry, are derived from the asymptotic free energy
per spin $f=-\lim_{N\to\infty}(\tilde{\beta}N)^{-1}\log Z_{\rm
s}$. Upon performing the trace over all geometries in
(\ref{eq:partsumslow}) one obtains, with
$\bsigma_i=(\sigma_i^1,\ldots,\sigma_i^n)$:
\begin{eqnarray}
Z_{\rm s} &=& \sum_{\bsigma^1\ldots \bsigma^n}\prod_{i<j}\left[ 1+
\frac{c}{N}e^{ \frac{\beta}{c}(\bxi_i\cdot \bxi_j) (\bsigma_i\cdot
\bsigma_j)}\right]
 \label{eq:Zprescaling}
\end{eqnarray}
In evaluating the free energy we make the usual `condensed'
ansatz: only a finite number $r$ of patterns will be structurally
correlated with the system state. The remaining $\alpha c-r$
patterns can be treated as frozen disorder, over which the free
energy may be averaged. For the result we write $[f]_{\rm dis}$.
In this paper we work within the connectivity scaling regime of
extreme dilution, where
$\lim_{N\to\infty}c/N=\lim_{N\to\infty}c^{-1}=0$. Now one obtains
\begin{eqnarray}
\hspace*{-15mm} -\tilde{\beta}[f]_{\rm dis}&=&
\lim_{N\to\infty}\frac{1}{N} \log
\sum_{\bsigma^1\ldots \bsigma^n}e^{ \frac{\beta}{2N}\sum_{i
j}(\bsigma_i\cdot \bsigma_j)\sum_{\mu\leq r}\xi^\mu_i
\xi^\mu_j+\frac{\alpha \beta^2}{4N} \sum_{ij}(\bsigma_i\cdot
\bsigma_j)^2+\order(\frac{N}{c})}
\end{eqnarray}
(modulo irrelevant additive constants). We define the familiar
 pattern and state overlaps
 $m_{\alpha \mu}(\bsigma)=N^{-1}\sum_i \xi_i^\alpha\sigma_i$ and  $q_{\alpha\beta}(\{\bsigma\})=N^{-1}\sum_i
\sigma_i^\alpha \sigma_i^\beta$. They are introduced via
appropriate $\delta$-distributions, so that the spin traces can be
carried out. This results in the usual type of steepest descent
expression for $[f]_{\rm dis}$ (again modulo constants):
\begin{eqnarray}
 [f]_{\rm dis}&=& {\rm extr}_{\{m_{\alpha\mu},q_{\alpha\beta}\}}\left\{
 \frac{\alpha\beta}{4n}
\sum_{\alpha\neq \beta}q^2_{\alpha\beta}+ \frac{1}{2n}
\sum_{\alpha}\sum_{\mu\leq r}m_{\alpha\mu}^2 \right.
\\
&&\left.
 -\frac{1}{n\beta}\bigbra\log
 \sum_{\sigma_1\ldots \sigma_n}e^{\beta\sum_{\alpha}\sum_{\mu\leq
r}m_{\alpha\mu}\xi_\mu\sigma_\alpha +\frac{1}{2}\alpha \beta^2
\sum_{\alpha\neq
\beta}q_{\alpha\beta}\sigma_\alpha\sigma_\beta}\bigket_\bxi
\right\}
\end{eqnarray}
where $\bra
g(\bxi)\ket_\bxi=2^{-r}\sum_{\bxi\in\{-1,1\}^r}g(\bxi)$. The
parameter $c$ represents the ensemble averaged connectivity. This
follows upon adding suitable generating fields to the slow
Hamiltonian: $H_{\rm s}(\bc)\to H_{\rm
s}(\bc)+\frac{2\lambda}{c}\sum_{i<j}c_{ij}$. Repeating  the above
calculation with the added fields shows that
$\lim_{N\to\infty}\frac{2}{Nc}\sum_{i<j}\overline{c_{ij}}=\lim_{\lambda\to
0}\frac{\partial [f]_{\rm dis}}{\partial \lambda}=1$, which proves
our claim.

 We next
make the replica-symmetric (RS) ansatz for the physical
saddle-point: $m_{\alpha\mu} = m_\mu$ for all $(\alpha,\mu)$ and
$q_{\alpha \beta} =q+\delta_{\alpha\beta}(1- q)$ for all
$(\alpha,\beta)$, keeping in mind that the replica dimension $n$
can take any non-negative value:
\begin{eqnarray}
 [f]^{\rm RS}_{\rm dis}&=& {\rm extr}_{\{m_\mu,q\}}\left\{
 \frac{1}{2}\sum_{\mu\leq r}m_{\mu}^2+
 \frac{1}{4}\alpha\beta[
  2q+(n-1) q^2]  \right.\nonumber
\\
&&\left.\hspace*{10mm}
 -\frac{1}{n\beta}\bra~\log\int\!Dz~ \cosh^n[
 \beta(\sum_{\mu\leq
r}m_{\mu}\xi_\mu + z\sqrt{\alpha q})] \ket_\bxi \right\}
\end{eqnarray}
Variation of $\{m_\mu,q\}$  gives the saddle-point equations for
our RS order parameters, with the short-hand $\Xi =
\beta(\sum_{\mu\leq r} m_\mu \xi_\mu + z\sqrt{\alpha q})$, which
are of the familiar form
\begin{eqnarray}
m_\mu
 &=& \bigbra \xi_\mu
\frac{\int\!Dz~\tanh(\Xi) \cosh^n(\Xi) }{
 \int\!Dz~\cosh^n(\Xi)} \bigket_{\!\bxi}
 \label{eq:RSm}
\\
q
  &=&
  \bigbra
\frac{ \int\!Dz~\tanh^2(\Xi)\cosh^n(\Xi) } { \int\!Dz~
\cosh^n(\Xi)}
 \bigket_{\!\bxi}
 \label{eq:RSq}
\end{eqnarray}
The physical meaning of the RS order parameters is
$m_\mu=\lim_{N\to\infty}N^{-1}\sum_i\overline{\bra \xi_i^\mu
\sigma_i\ket}$ and $q=\lim_{N\to\infty}N^{-1}\sum_i
\overline{\bra\sigma_i\ket^2}$,  as usual.

\subsection{Phase transitions and phase diagrams}

If one simplifies matters further by assuming only one pattern to
be condensed, i.e.
 $m_\mu=m\delta_{\mu,1}$, then equations
 (\ref{eq:RSm},\ref{eq:RSq}) reduce to
\begin{eqnarray}
m
 &=& \frac{\int\!Dz~\tanh[\beta(m + z\sqrt{\alpha q})] \cosh^n[\beta( m
 + z\sqrt{\alpha q})] }{
 \int\!Dz~\cosh^n[\beta(m+ z\sqrt{\alpha q})]}
 \label{eq:RSmsimple}
\\
q
  &=&
\frac{ \int\!Dz~\tanh^2[\beta( m+ z\sqrt{\alpha
q})]\cosh^n[\beta(m+ z\sqrt{\alpha q})] } { \int\!Dz~
\cosh^n[\beta(m + z\sqrt{\alpha q})]}
 \label{eq:RSqsimple}
\end{eqnarray}
 These are recognized to be identical to
 those of the finite $n$ model studied in \cite{Sherrington} if we re-define the parameters in the latter
 according to
\begin{eqnarray}
J^{(1)} m/k ~\to~ m ~~~~~~~~J^{(2)} q/k ~\to~ \alpha\beta q
\end{eqnarray}
This makes sense, since the $n\to 0$ limit of our present model
(i.e. the symmetrically extremely diluted Hopfield model with
quenched random connectivity \cite{WS}) is known to map onto the
$n\to 0$ limit of \cite{Sherrington} (i.e. the SK model
\cite{SK}). Clearly one finds simplified equations for the special
dimension values  $n=1$ (equivalent to having annealed geometry)
and $n=2$, where the Gaussian integrals can be done. For instance,
at $n=1$ the equation for $m$ reduces to $m = \tanh(\beta m)$,
whereas
 for $n=2$ one finds
\begin{eqnarray}
n=2:~~~~~ m=\frac{\sinh(2\beta m)}{\cosh(2\beta
m)+e^{-2\alpha\beta^2 q}} ~~~~~ q = \frac{\cosh(2\beta m) -
e^{-2\alpha \beta^2 q}} {\cosh(2\beta m) + e^{-2\alpha \beta^2 q}}
\end{eqnarray}
Our RS equations admit three phases: a paramagnetic phase (P) with
$m=q=0$, a recall phase (R) where $m\neq 0$ and $q>0$, and a
spin-glass phase (SG) where $m=0$ but $q>0$. Since deriving the RS
phase transitions has been reduced to appropriate translation of
the results found in \cite{Sherrington}, we will here simply
mention the outcome:
\begin{figure}[t]
\vspace*{-4mm}\hspace*{30mm} \setlength{\unitlength}{0.8mm}
\begin{picture}(100,100)
\put(15,-0){\epsfysize=95\unitlength\epsfbox{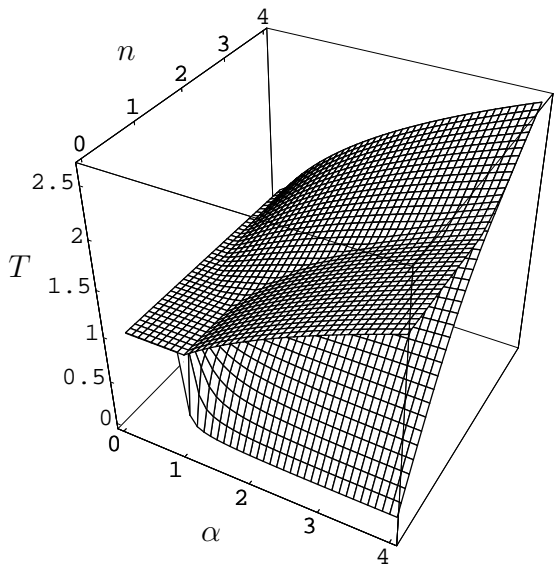}}
 \put(44,5){$\alpha$}  \put(12,49){$T$} \put(30,85){$n$}
\end{picture}
\caption{RS phase diagram in the space of control parameters. We
show the critical temperature(s)
 as surface(s) over the $(\alpha,n)$
plane. The high temperature phase is paramagnetic (P). At low
temperature we find the retrieval  phase R. For $(\alpha,n)$
values with two critical temperatures, the latter define the
boundaries of a spin-glass phase SG. The P$\to$R transitions are
second order for $\alpha<\frac{1}{3n-2}$, and first order
elsewhere. The P$\to$SG transitions are second order for $n< 2$
and first order elsewhere. For large $\alpha$ the SG$\to$R
transitions become second order, but for small $\alpha$ they are
first order.} \label{fig:phased}
\end{figure}
\begin{itemize}
\item
For sufficiently small $\alpha$ one will find a P$\to$R transition
at a finite temperature. For $\alpha<\alpha_c = \frac{1}{3n-2}$
this transition is second order, and occurs at $T_{\rm R}=1$.
\item
For larger $\alpha$, lowering temperature will lead first to a
P$\to$SG transition, followed at some yet lower temperature by a
SG$\to$R transition\footnote{For small values of $n$, the latter
SG$\to$R transition is expected to disappear when replica symmetry
breaking is taken into account.}. For $n\leq 2$ the P$\to$SG
transition is second order, and occurs for $T_{\rm
SG}=\sqrt{\alpha}$.
\item
The SG$\to$R transition is second order for $\alpha\to \infty$,
where its transition temperature tends to $T_{c}=n$, but first
order for sufficiently small $\alpha$.
\item
The effects of increasing the replica dimension $n$ are (i) a
reduction of the size in the phase diagram of the SG phase, and
(ii) a change of the orders of the P$\to$R and P$\to$SG
transitions, from second order (for small $n$) to first order (for
large $n$).
\end{itemize}
\begin{figure}[t]
\vspace*{-4mm} \setlength{\unitlength}{0.65mm}
\begin{picture}(200,100)
 \put(0,100){\epsfig{file=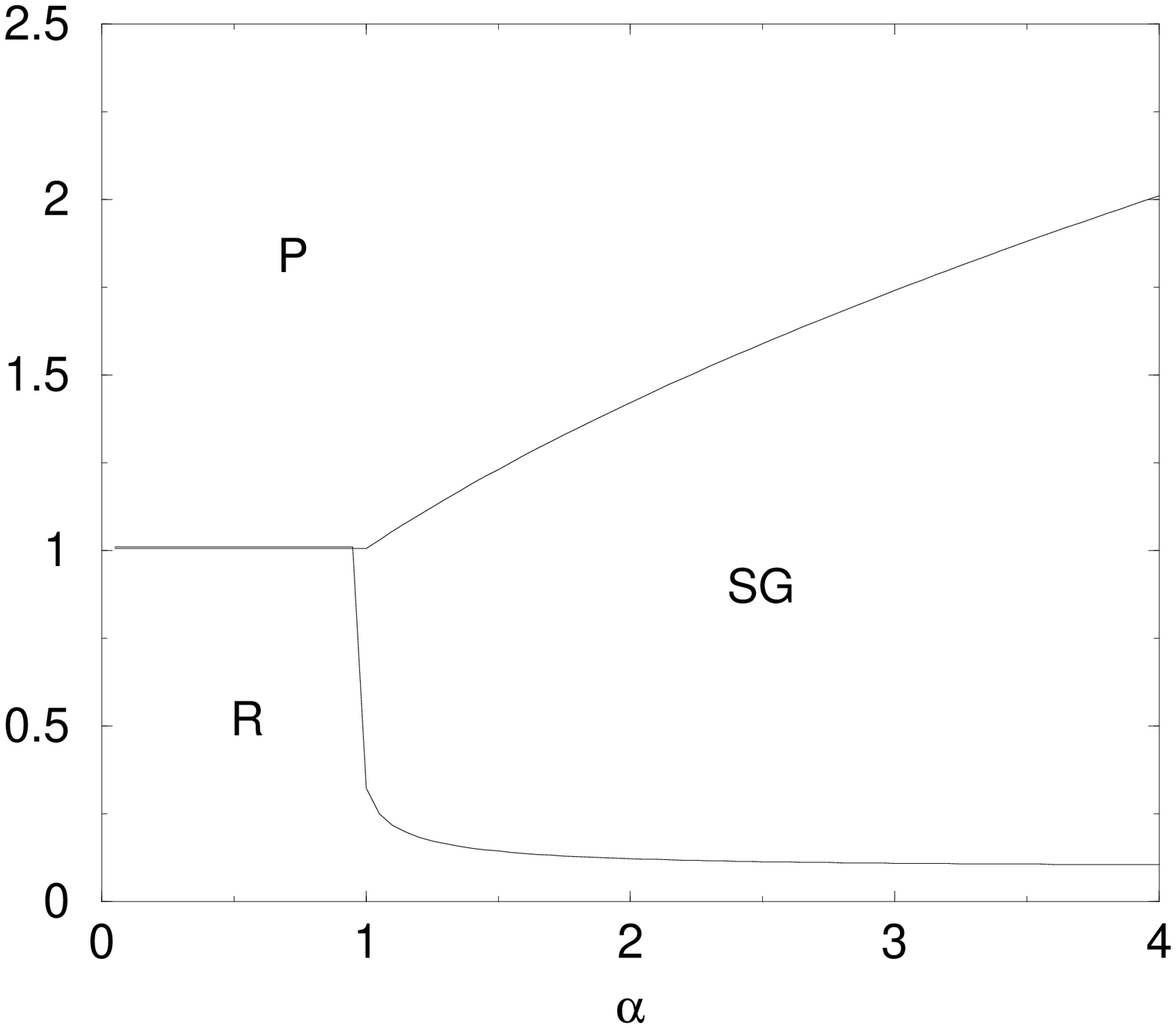, angle=270,
width=115\unitlength}}
 \put(126,100){\epsfig{file=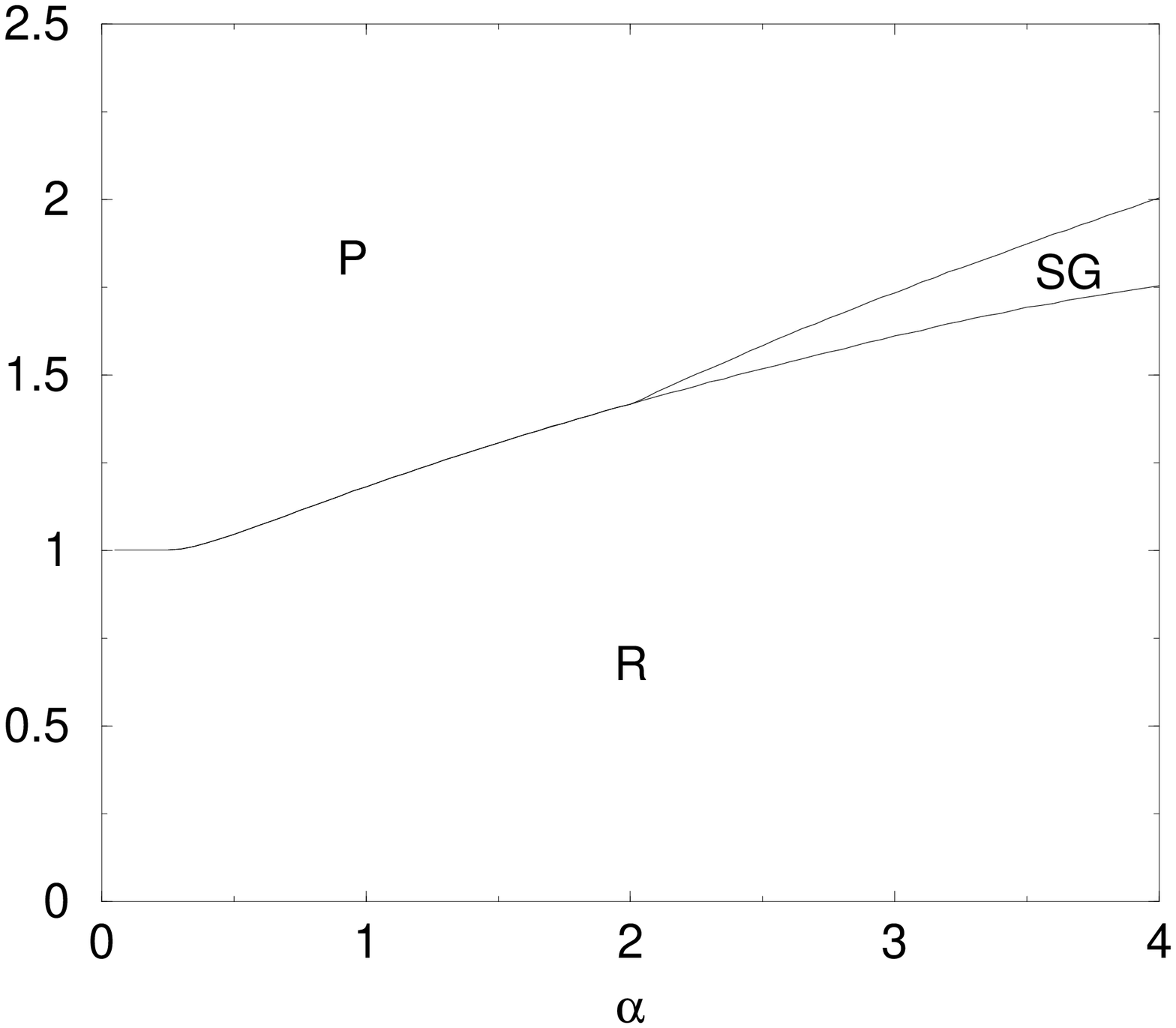,
angle=270, width=115\unitlength}} \put(-1,47){\small $T$}
\put(125,47){\small $T$} \put(85,80){\small $n=0.1$}
\put(216,80){\small $n=2$}
\end{picture}
\vspace*{4mm} \caption{Intersections of the phase diagram shown in
figure \ref{fig:phased},  at $n=0.1$ (left) and  $n=2$ (right). We
have paramagnetic (P), recall (R) and spin-glass (SG) phases. We
note that there is no critical value for $\alpha$ above which
recall is no longer possible. Instead the SG$\to$R transition line
will approach the line $T=n$ for large $\alpha$. Since RSB
phenomena appear to be  confined to $n<1$ (see below), this is not
an artifact of the RS assumption. For large $n$ all phase
transitions ultimately become first order.} \label{fig:intersect}
\end{figure}
Numerical solution of equations
(\ref{eq:RSmsimple},\ref{eq:RSqsimple}) leads to the RS phase
diagram drawn in figure \ref{fig:phased}. Figure
\ref{fig:intersect} shows intersections of this diagram in the
planes of constant replica dimension $n=0.1$ and $n=2$. All
transitions discussed and drawn above refer to bifurcations of
locally stable solutions, since for recurrent neural networks the
time-scales where thermodynamic stability would become an issue
are in practice never reached.

\begin{figure}[t]
\vspace*{5mm} \hspace*{40mm}\setlength{\unitlength}{0.65mm}
\begin{picture}(200,100)
 \put(0,100){\epsfig{file=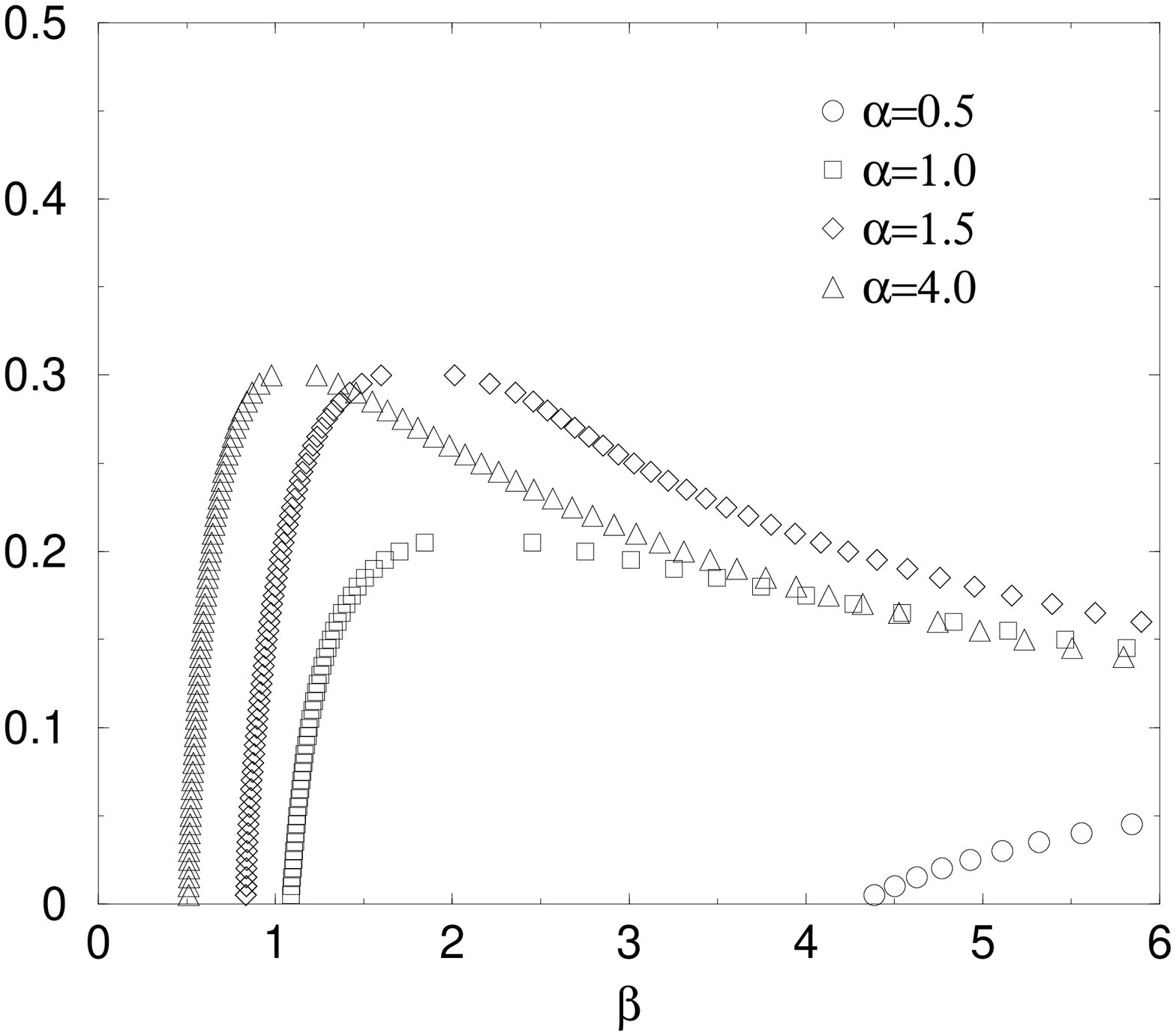, angle=270,
width=115\unitlength}} \put(-5,53){$n_c$}
\end{picture}
\vspace*{-7mm}
 \caption{Location of the AT instability $n_c$, shown as a function of the inverse temperature
 $\beta=T^{-1}$ ,and for a number of different storage ratios $\alpha$.   Replica symmetry
breaking is seen to be limited to values of $n$ below $0.32$. We
also  note the non-monotonic dependence on temperature of the
critical dimension $n_c$ for fixed $\alpha$. } \label{fig:AT}
\end{figure}

We finally turn to replica symmetry breaking (RSB). The location
in our phase diagram of second order RSB phase transition follow
upon inspection of the eigenvalues of the Hessian. Since our model
can be mapped onto the nonzero-$n$ SK-model  \cite{Sherrington},
 we can read off the
eigenvalues from \cite{AT}. The dangerous eigenvalue $\lambda_{\rm
RSB}$ is the one associated with the so-called replicon mode:
\begin{eqnarray}
\lambda_{\rm RSB} &=& \alpha \beta^2\left[1-\alpha
\beta^2[1-2q+h(m,q)]\right] \\
 h(m,q)&=& \frac{\int\! Dz~
\tanh^4[\beta( m+ z\sqrt{\alpha q})]\cosh^n[\beta(m+ z\sqrt{\alpha
q})]}{\int\! Dz~ \cosh^n[\beta(m+ z\sqrt{\alpha q})]}
\label{eq:hmqeq}
\end{eqnarray}
Replica symmetry is stable only if $\lambda_{\rm RSB}>0$. For each
combination $(\alpha,T)$ one finds a critical value
$n_c(\alpha,T)$ (the AT line) below which replica symmetry is
unstable. Examples of the the behaviour of $n_c(\alpha,T)$ are
shown in figure \ref{fig:AT}. Replica symmetry breaking is found
to occur only for $n<0.32$.

 Compared to
diluted neural network models with static random geometry, the
main effect of introducing dynamic geometry (with the present
Glauber dynamics, aimed at reducing frustration) is to reduce the
spin-glass phase in favour of the recall phase. The geometry
adjusts itself autonomously in order to retrieve the condensed
pattern optimally, to such an extent that for sufficiently low
temperature there is no upper limit on the storage ratio (provided
we do not leave the `extreme dilution' scaling regime
$\lim_{N\to\infty}c/N=\lim_{N\to \infty}c^{-1}=0$).
 This then raises the question of
whether the other (non-condensed) patterns can be retrieved at all
after the geometry has been taylored  to the recall of one
specific condensed pattern. This is investigated in section
\ref{sec:tausys}.

\section{Fraction of misaligned spins}

We expect the observed improvement of  retrieval performance due
to the slow geometry dynamics to be reflected in a reduction  with
increasing $n$ of the fraction of frustrated bonds in the system.
To verify this we calculate a different but similar quantity: the
fraction $\phi$ of misaligned spins, i.e. those where $\sigma_i$
and  local field $h_i$ have opposite sign:
\be
\phi=\lim_{N\to\infty}\frac{1}{N}\sum_i[\overline{\bra\theta[
-\sigma_i\sum_j \frac{c_{ij}}{c} \sum_\mu \xi_i^\mu \xi_j^\mu
\sigma_j]\ket}]_{\rm dis} \label{eq:define_phi} \ee To calculate
this object one could introduce further replicas, but here we
follow an alternative route: we solve our model first for finite
$c$, in which case joint replicated spin-field distributions (in
terms of which $\phi$ can be expressed) become the natural  order
parameters, followed by taking the limit $c\to\infty$.

\subsection{Calculation of the joint spin-field distribution}

To do so we have to adapt and generalize  the calculation in e.g.
\cite{ThomsenThorpe} by first introducing the $2^{p}$ so-called
sub-lattices \cite{Hemmen}, with
$\bxi_i=(\xi_i^1,\ldots,\xi_i^p)$:
\begin{equation}
I_{\bxi} = \{i| \bxi_i = \bxi \} ~~~~~~~~~~~~~~ p_\bxi =
|I_\bxi|/N
\end{equation}
Since $c$ is assumed finite, so is $p=\alpha c$. We write
$\sum_{\bxi}p_\bxi \Phi(\bxi)=\bra\Phi(\bxi)\ket_\bxi$; for
randomly drawn patterns $\lim_{N\to\infty}p_{\bxi}=2^{-p}$. For
finite $c$ our analysis will start to resemble that in e.g.
\cite{WemmenCoolen}. In each sublattice we may define a joint
distribution for replicated spins and fields, and (with a modest
amount of foresight) conjugate fields:
\begin{eqnarray}
P_\bxi(\bsigma,\bh,\hat{\bh})&=&|I_\bxi|^{-1}\sum_{i\in
I_\bxi}\delta_{\bsigma,\bsigma_i}\delta[\bh-\bh_i(\{\bsigma\})]\delta[\hat{\bh}-\hat{\bh}_i(\{\bsigma\})]
\label{eq:spin_field_dist}
 \end{eqnarray}
 where $\bsigma\in\{-1,1\}^n$,
$\bh,\hat{\bh}\in\R^n$, and $h_i^\alpha(\{\bsigma\})=\sum_j
\frac{c_{ij}}{c}(\bxi_i\cdot \bxi_j) \sigma^\alpha_j$. In
evaluating the free energy per spin we write the fast Hamiltonian
in terms of replicated fields, and introduce
(\ref{eq:spin_field_dist}) by inserting suitable integrals over
$\delta$-functions. This is done first only for discrete values of
$\bh$, with the continuum limit (converting integrals into path
integrals) to be taken after the thermodynamic limit. We
abbreviate $\{dP
d\hat{P}\}=\prod_{\bxi,\bsigma,\bh,\hat{\bh}}[dP_{\bxi}(\bsigma,\bh,\hat{\bh})d\hat{P}_{\bxi}(\bsigma,\bh,\hat{\bh})]$
 and find
\begin{eqnarray}
\hspace*{-20mm}
 -\tilde{\beta} f &=& \lim_{N\to\infty}\frac{1}{N} \log
\sum_{\bc}\sum_{\bsigma^1\ldots \bsigma^n} e^{\frac{\beta}{2}
\sum_{i}\bsigma_i\cdot \bh_i(\{\bsigma\})-\log(\frac{N}{c})
\sum_{i<j}c_{ij}} \nonumber
\\
\hspace*{-20mm} &=& \lim_{N\to\infty}\frac{1}{N} \log \int\!\{ dP
d\hat{P}\}e^{N\bra \sum_{\bsigma}\int\!d\bh d\hat{\bh}
~P_{\bxi}(\bsigma,\bh,\hat{\bh})\left[ i
\hat{P}_{\bxi}(\bsigma,\bh,\hat{\bh}) + \frac{1}{2}\beta
(\bsigma\cdot\bh)\right]\ket_{\bxi}} \nonumber
\\
\hspace*{-20mm} &&\times \int\!\prod_i\left[\frac{d\bh_i
d\hat{\bh}_i}{(2\pi)^n}e^{i\hat{\bh}_i\cdot\bh_i}\right]\sum_{\bsigma^1\ldots
\bsigma^n} e^{-i\sum_{\bxi}\sum_{i\in I_\bxi}
\hat{P}_{\bxi}(\bsigma_i,\bh_i,\hat{\bh}_i)} \nonumber
\\
\hspace*{-20mm} &&\times  \prod_{i<j}\left[ 1+ \frac{c}{N}
 e^{- \frac{i}{c} (\bxi_i\cdot\bxi_j)[(\hat{\bh}_i \cdot \bsigma_j) +(\hat{\bh}_j \cdot \bsigma_i)]}
\right] \nonumber
\\
\hspace*{-20mm} &=& \lim_{N\to\infty}\frac{1}{N} \log \int\!\{ dP
d\hat{P}\}e^{N\bra \sum_{\bsigma}\int\!d\bh d\hat{\bh}
~P_{\bxi}(\bsigma,\bh,\hat{\bh})\left[ i
\hat{P}_{\bxi}(\bsigma,\bh,\hat{\bh}) + \frac{1}{2}\beta
(\bsigma\cdot\bh)\right]\ket_{\bxi}} \nonumber
\\
\hspace*{-20mm} && \times e^{
\frac{c}{2}N\bra\bra\sum_{\bsigma\bsigma^\prime}\int\!d\bh
d\bh^\prime d\hat{\bh}d\hat{\bh}^\prime
~P_{\bxi}(\bsigma,\bh,\hat{\bh})P_{\bxi^\prime}(\bsigma^\prime,\bh^\prime,\hat{\bh}^\prime)
 e^{- \frac{i}{c} (\bxi\cdot\bxi^\prime)[(\hat{\bh} \cdot \bsigma^\prime) +(\hat{\bh}^\prime \cdot
 \bsigma)]}\ket_\bxi\ket_{\bxi^\prime}}
 \nonumber
\\
\hspace*{-20mm} &&\times \int\!\prod_i\left[\frac{d\bh_i
d\hat{\bh}_i}{(2\pi)^n}e^{i\hat{\bh}_i\cdot\bh_i}\right]\sum_{\bsigma^1\ldots
\bsigma^n} e^{-i\sum_{\bxi}\sum_{i\in I_\bxi}
\hat{P}_{\bxi}(\bsigma_i,\bh_i,\hat{\bh}_i)} \nonumber
\\
\hspace*{-20mm} &=& \lim_{N\to\infty}\frac{1}{N} \log \int\!\{ dP
d\hat{P}\}e^{N\bra \sum_{\bsigma}\int\!d\bh d\hat{\bh}
~P_{\bxi}(\bsigma,\bh,\hat{\bh})\left[ i
\hat{P}_{\bxi}(\bsigma,\bh,\hat{\bh}) + \frac{1}{2}\beta
(\bsigma\cdot\bh)\right]\ket_{\bxi}} \nonumber
\\
\hspace*{-20mm} && \times e^{
\frac{c}{2}N\bra\bra\sum_{\bsigma\bsigma^\prime}\int\!d\bh
d\bh^\prime d\hat{\bh}d\hat{\bh}^\prime
~P_{\bxi}(\bsigma,\bh,\hat{\bh})P_{\bxi^\prime}(\bsigma^\prime,\bh^\prime,\hat{\bh}^\prime)
 e^{- \frac{i}{c} (\bxi\cdot\bxi^\prime)[(\hat{\bh} \cdot \bsigma^\prime) +(\hat{\bh}^\prime \cdot
 \bsigma)]}\ket_\bxi\ket_{\bxi^\prime}}
 \nonumber
\\
\hspace*{-20mm} &&\times e^{N\bigbra\log\left\{
\int\!\left[\frac{d\bh
d\hat{\bh}}{(2\pi)^n}e^{i\hat{\bh}\cdot\bh}\right]\sum_{\bsigma\in\{-1,1\}^n}
e^{-i \hat{P}_{\bxi}(\bsigma,\bh,\hat{\bh})}
\right\}\bigket_{\bxi}}
 \nonumber
 \\
 \hspace*{-20mm}
&=& {\rm extr}_{\{ P,\hat{P}\}}\left\{ \bra
\sum_{\bsigma}\int\!d\bh d\hat{\bh}
~P_{\bxi}(\bsigma,\bh,\hat{\bh})\left[ i
\hat{P}_{\bxi}(\bsigma,\bh,\hat{\bh}) + \frac{1}{2}\beta
(\bsigma\cdot\bh)\right]\ket_{\bxi} \right. \nonumber
\\
\hspace*{-20mm} &&\left.
+\frac{1}{2}c\bra\bra\sum_{\bsigma\bsigma^\prime}\int\!d\bh
d\bh^\prime d\hat{\bh}d\hat{\bh}^\prime
~P_{\bxi}(\bsigma,\bh,\hat{\bh})P_{\bxi^\prime}(\bsigma^\prime,\bh^\prime,\hat{\bh}^\prime)
 e^{- \frac{i}{c} (\bxi\cdot\bxi^\prime)[(\hat{\bh} \cdot \bsigma^\prime) +(\hat{\bh}^\prime \cdot
 \bsigma)]}\ket_\bxi\ket_{\bxi^\prime}
 \right.\nonumber
\\
\hspace*{-25mm} &&\left. + \bigbra\log\left\{
\int\!\left[\frac{d\bh
d\hat{\bh}}{(2\pi)^n}e^{i\hat{\bh}\cdot\bh}\right]\sum_{\bsigma\in\{-1,1\}^n}
e^{-i \hat{P}_{\bxi}(\bsigma,\bh,\hat{\bh})}
\right\}\bigket_{\bxi}
 \right\}
\end{eqnarray}
Extremization with respect to $P_{\bxi}(\bsigma,\bh,\hat{\bh})$
and $\hat{P}_{\bxi}(\bsigma,\bh,\hat{\bh})$ gives the following
two saddle-point equations:
\begin{eqnarray}
\hspace*{-10mm} \hat{P}_{\bxi}(\bsigma,\bh,\hat{\bh})&=& ic
\bra\sum_{\bsigma^\prime}\int\! d\hat{\bh}^\prime
~P_{\bxi^\prime}(\bsigma^\prime,\hat{\bh}^\prime)
 e^{- \frac{i}{c} (\bxi\cdot\bxi^\prime)[(\hat{\bh} \cdot \bsigma^\prime) +(\hat{\bh}^\prime \cdot
 \bsigma)]}\ket_{\bxi^\prime}
 +\frac{1}{2}i\beta
(\bsigma\cdot\bh) \label{eq:sp_Phat}
\\
\hspace*{-10mm} P_{\bxi}(\bsigma,\bh,\hat{\bh}) &=&
 \frac{e^{i\hat{\bh}\cdot\bh-i
 \hat{P}_{\bxi}(\bsigma,\bh,\hat{\bh})}}
 { \int\!d\bh^\prime
d\hat{\bh}^\prime~
e^{i\hat{\bh}^\prime\cdot\bh^\prime}\sum_{\bsigma^\prime\in\{-1,1\}^n}
e^{-i \hat{P}_{\bxi}(\bsigma^\prime,\bh^\prime,\hat{\bh}^\prime)}}
\label{eq:sp_P}
\end{eqnarray}
Insertion of (\ref{eq:sp_Phat}) into (\ref{eq:sp_P}) gives a
saddle-point equation in terms of $P$ only:
\begin{eqnarray}
\hspace*{-10mm} P_{\bxi}(\bsigma,\bh,\hat{\bh}) &=& Z_{\bxi}^{-1}
 e^{i\hat{\bh}\cdot\bh +\frac{1}{2}\beta
(\bsigma\cdot\bh)+c \bra\sum_{\bsigma^\prime}\int\!
d\hat{\bh}^\prime
~P_{\bxi^\prime}(\bsigma^\prime,\hat{\bh}^\prime)
 e^{- \frac{i}{c} (\bxi\cdot\bxi^\prime)[(\hat{\bh} \cdot \bsigma^\prime) +(\hat{\bh}^\prime \cdot
 \bsigma)]}\ket_{\bxi^\prime}
 }\label{eq:final_P}
\end{eqnarray}
with $Z_{\bxi}$ denoting a normalization constant. According to
(\ref{eq:spin_field_dist}), the physical meaning of the
saddle-point is
\begin{eqnarray}
P_\bxi(\bsigma,\bh,\hat{\bh})&=&\lim_{N\to\infty}\frac{1}{|I_\bxi|}\sum_{i\in
I_\bxi}\overline{\bra
\delta_{\bsigma,\bsigma_i}\delta[\bh-\bh_i(\{\bsigma\})]\delta[\hat{\bh}-\hat{\bh}_i(\{\bsigma\})]\ket}
\label{eq:physical_meaning_of_P}
 \end{eqnarray}
 We next make the one
pattern condensed ansatz (this is not essential for being able to
proceed, but will simplify and compactify our derivation
significantly), which here implies
$P_{\bxi}(\bsigma,\bh,\hat{\bh})=P_{\xi_1}(\bsigma,\bh,\hat{\bh})$,
and we send $c\to\infty$. As a result
$(\bxi\cdot\bxi^\prime)/\sqrt{c}=\xi_1\xi_1^\prime/\sqrt{c}
+\sqrt{\alpha} z$ where $z$ is a zero-average unit-variance
Gaussian variable, and
\begin{eqnarray}
\hspace*{-21mm} P_{\xi}(\bsigma,\bh,\hat{\bh})
 &=& Z^{-1}_{\xi}
 e^{i\hat{\bh}\cdot\bh +\frac{1}{2}\beta
(\bsigma\cdot\bh)- \bra\sum_{\bsigma^\prime}\int\!
d\hat{\bh}^\prime ~P_{\xi^\prime}(\bsigma^\prime,\hat{\bh}^\prime)
 \left[i(\xi\xi^\prime)[(\hat{\bh} \cdot
\bsigma^\prime) +(\hat{\bh}^\prime \cdot
 \bsigma)]+
 \frac{\alpha}{2}[(\hat{\bh} \cdot \bsigma^\prime) +(\hat{\bh}^\prime \cdot
 \bsigma)]^2\right]
\ket_{\xi^\prime}
 } \nonumber
 \\
 \hspace*{-23mm}&&
 \label{eq:condensed_joint}
\end{eqnarray}
In the right-hand side of (\ref{eq:condensed_joint}) we are seen
to need only the following moments of our distributions
 (which include the previously
encountered $\{m_\alpha,q_{\alpha\beta}\}$):
\begin{eqnarray*}
m_\alpha&=\bra \xi
\sum_{\bsigma}\int\!d\hat{\bh}~P_{\xi}(\bsigma,\hat{\bh})\sigma_\alpha\ket_{\xi}
~~~~~~~~~ q_{\alpha\beta}&=\bra
\sum_{\bsigma}\int\!d\hat{\bh}~P_{\xi}(\bsigma,\hat{\bh})\sigma_\alpha
\sigma_\beta\ket_{\xi}\\ k_{\alpha}&=i\bra \xi
\sum_{\bsigma}\int\!d\hat{\bh}~P_{\xi}(\bsigma,\hat{\bh})\hat{h}_\alpha\ket_{\xi}~~~~~~~~
L_{\alpha\beta}&=\bra
\sum_{\bsigma}\int\!d\hat{\bh}~P_{\xi}(\bsigma,\hat{\bh})\hat{h}_\alpha
\hat{h}_\beta \ket_{\xi}\\ K_{\alpha\beta }&=i\bra
\sum_{\bsigma}\int\!d\hat{\bh}~P_{\xi}(\bsigma,\hat{\bh})\sigma_\alpha
\hat{h}_\beta \ket_{\xi}&
\end{eqnarray*}
Integration by parts over the fields in (\ref{eq:condensed_joint})
shows that $k_\alpha=-\frac{1}{2}\beta m_\alpha$,
$K_{\alpha\beta}=-\frac{1}{2}\beta q_{\alpha\beta}$, and
$L_{\alpha\beta}=-\frac{1}{4}\beta^2 q_{\alpha\beta}$. The
replicated joint spin-field distributions can now be written as
\begin{eqnarray}
\hspace*{-15mm}
 P_{\xi}(\bsigma,\bh) &=&  \frac{ e^{\xi\beta\bm\cdot\bsigma+
 \frac{1}{2}\alpha\beta^2\bsigma \cdot\bq\bsigma -\frac{1}{2\alpha}(\bh -
 \xi\bm -\alpha \beta \bq \bsigma)\bq^{-1}(\bh -
 \xi\bm -\alpha \beta \bq \bsigma)}
} {\sum_{\bsigma^\prime}\!\int\!d\bh^\prime~
e^{\xi\beta\bm\cdot\bsigma^\prime
 +\frac{1}{2}\alpha\beta^2\bsigma^\prime \cdot\bq\bsigma^\prime -\frac{1}{2\alpha}(\bh^\prime -
 \xi\bm -\alpha \beta \bq \bsigma^\prime)\bq^{-1}(\bh^\prime -
 \xi\bm -\alpha \beta \bq \bsigma^\prime)}
} \label{eq:replicated_spinfield}
\end{eqnarray}
with $\bm=\{m_\alpha\}$ and $\bq=\{q_{\alpha\beta}\}$.  The latter
obey the following familiar closed equations which in RS ansatz
lead one back to (\ref{eq:RSmsimple}), as they should:
\begin{eqnarray}
 m_\alpha &=&  \bigbra \xi \frac{\sum_{\bsigma} \sigma_\alpha e^{\xi\beta\bm\cdot\bsigma+
 \frac{1}{2}\alpha\beta^2\bsigma \cdot\bq\bsigma }
} {\sum_{\bsigma} e^{\xi\beta\bm\cdot\bsigma
 +\frac{1}{2}\alpha\beta^2\bsigma \cdot\bq\bsigma}
}\bigket_{\!\!\xi}
\\
 q_{\alpha\beta} &=&  \bigbra\frac{\sum_{\bsigma}\sigma_\alpha\sigma_\beta e^{\xi\beta\bm\cdot\bsigma+
 \frac{1}{2}\alpha\beta^2\bsigma \cdot\bq\bsigma }
} {\sum_{\bsigma} e^{\xi\beta\bm\cdot\bsigma
 +\frac{1}{2}\alpha\beta^2\bsigma\cdot\bq\bsigma}
}\bigket_{\!\!\xi}
\end{eqnarray}

\subsection{Fraction of mis-aligned spins in RS ansatz}

\begin{figure}[t]
\vspace*{-8mm} \hspace*{5mm}
 \setlength{\unitlength}{0.58mm}
\begin{picture}(270,100)
 \put(0,100){\epsfig{file=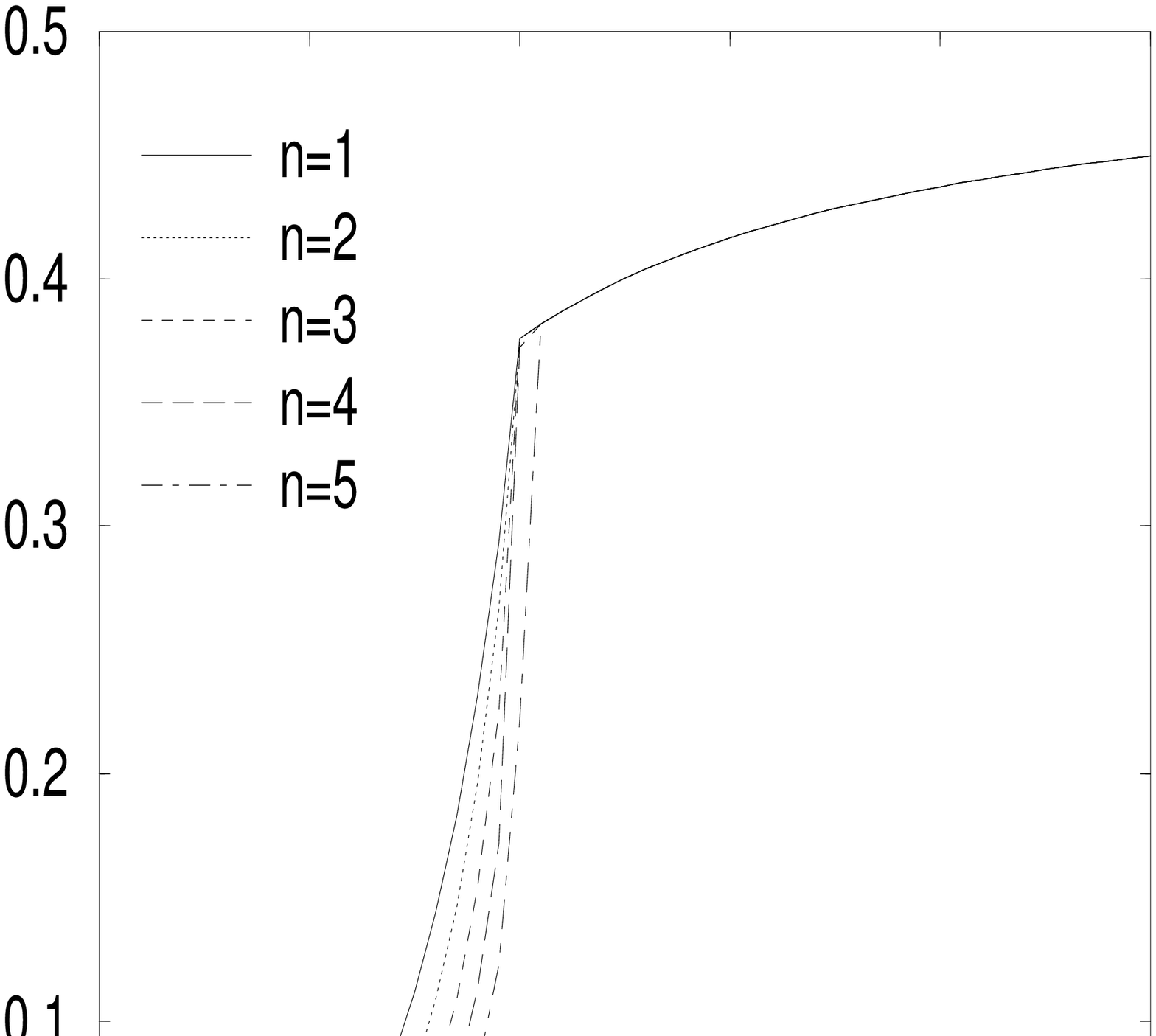, angle=270, width=90\unitlength}}
 \put(85,100){\epsfig{file=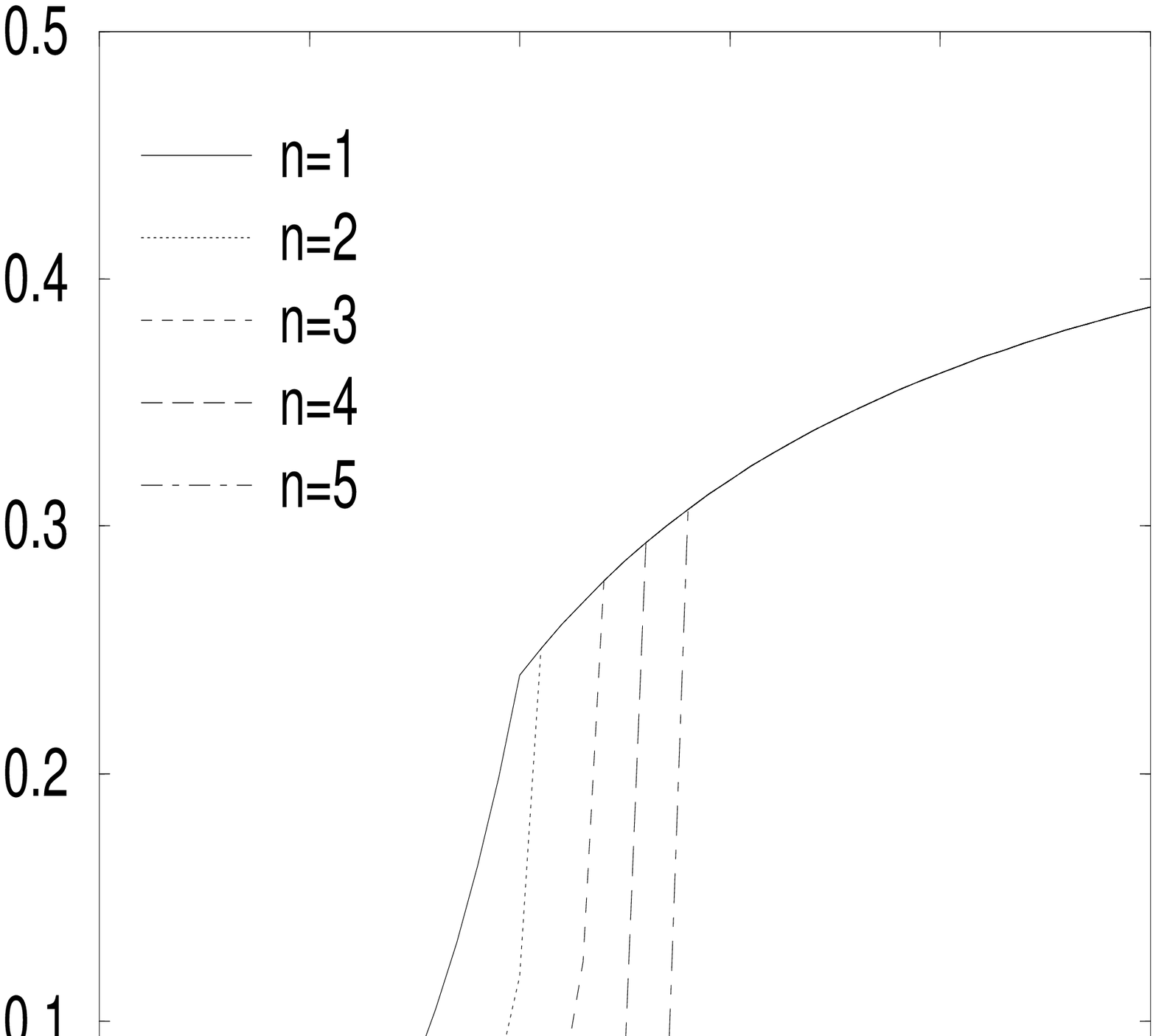, angle=270, width=90\unitlength}}
 \put(170,100){\epsfig{file=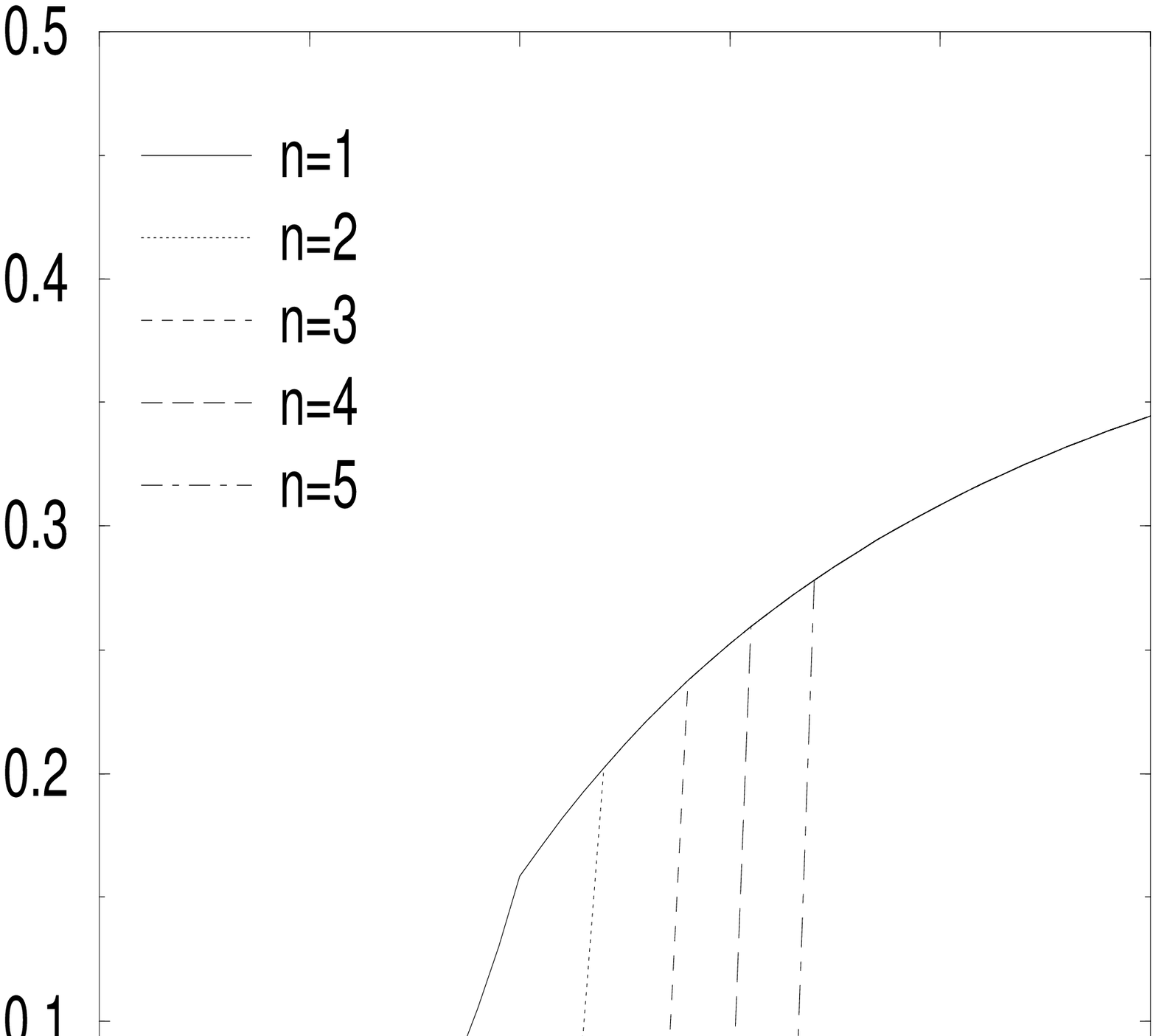, angle=270, width=90\unitlength}}
 \put(-9,35){\small $\phi_{\rm RS}$}
\put(47,-24){\small $T$} \put(132,-24){\small $T$}
\put(217,-24){\small $T$} \put(40,72){\footnotesize $\alpha=0.1$}
\put(125,72){\footnotesize $\alpha=0.5$}
\put(210,72){\footnotesize $\alpha=1.0$}
\end{picture}
\vspace*{10mm} \caption{The fraction $\phi_{\rm RS}$ of misaligned
spins as function of temperature, in RS ansatz.  for integer
values of $n$ between $1$ and $5$, and $\alpha=0.1$, $\alpha=0.5$
and $\alpha=1$ respectively, as a function of temperature. The
degree of alignment of spins to their local fields increases with
$n$ outside the paramagnetic phase. In the paramagnetic phase
there is no dependence on $n$. One sees clearly the effect of the
first order phase transitions, at $\alpha=0.1$ only for $n=5$, and
for $\alpha=0.5$ and $\alpha=1$ for all $n \geq 2$.}
\label{fig:misaligned}
\end{figure}

The fraction $\phi$ defined in (\ref{eq:define_phi}) can be
written as $\phi=\bra \phi_\xi\ket_\xi$, where the sublattice
fractions $\phi_\xi$ are expressed in terms of the replicated
distributions (\ref{eq:replicated_spinfield}) in the following
way:
\begin{eqnarray}
\hspace*{-23mm} \phi_\xi&=&\frac{1}{2}-\frac{1}{2}
\sum_{\bsigma}\int\!d\bh~P_\xi(\bsigma,\bh)\frac{1}{n}\sum_\gamma
\sigma_\gamma \sgn[ h_\gamma] \nonumber
\\
\hspace*{-23mm} &=&\frac{1}{2}-\frac{1}{2n}\sum_\gamma
\frac{\sum_{\bsigma}\sigma_\gamma e^{\beta\bm\cdot\bsigma+
 \frac{1}{2}\alpha\beta^2\bsigma \cdot\bq\bsigma}\int\!d\bx~
\sgn[m_\gamma +\alpha \beta \sum_\beta q_{\gamma\beta}
\sigma_\beta +\sqrt{\alpha} x_\gamma]
e^{-\frac{1}{2}\bx\cdot\bq^{-1}\bx} }
{\sum_{\bsigma}e^{\beta\bm\cdot\bsigma
 +\frac{1}{2}\alpha\beta^2\bsigma\cdot\bq\bsigma}\int\!d\bx~ e^{-\frac{1}{2}\bx\cdot\bq^{-1}\bx} }
 \nonumber
 \\
 \hspace*{-23mm}&&
\end{eqnarray}
We see that the $\phi_1=\phi_{-1}$. At this stage we make the RS
ansatz, putting $m_\alpha=m$ and
$q_{\alpha\beta}=q+\delta_{\alpha\beta}(1-q)$, which results in
\begin{eqnarray}
\hspace*{-20mm} \phi_{\rm RS}&=& \frac{1}{2}-\frac{1}{2}
\frac{\int\!Dz \sum_{\bsigma}\sigma_1 e^{\beta
\sum_\alpha\sigma_\alpha(m+
  z\sqrt{\alpha q})}\int\!Dx~
\sgn[m +\alpha \beta (\sigma_1+ q\sum_{\beta>1} \sigma_\beta)
+\sqrt{\alpha} x]} { \int\!Dz \sum_{\bsigma} e^{\beta
\sum_\alpha\sigma_\alpha(m+
  z\sqrt{\alpha q})}}
 \nonumber\\
\hspace*{-20mm}
 &=&
\frac{1}{2}-\frac{1}{2} \frac{\int\!DxDyDz \sum_{\bsigma}\sigma_1
e^{\beta \sum_\alpha\sigma_\alpha(m+
  z\sqrt{\alpha
  q})+(x-iy)\sqrt{\alpha}\beta(\sigma_1+q\sum_{\alpha>1}\sigma_\alpha)}
\sgn[m +\sqrt{\alpha} x]} { \int\!Dz \left[2\cosh[\beta(m+
  z\sqrt{\alpha q})]\right]^n}
  \nonumber
  \end{eqnarray}
We carry out the spin summations over $\sigma_{\alpha}$ with
$\alpha>1$. A shift in the complex plane for the variable $z$ in
the numerator, $z\to z-\sqrt{q}(x-iy)$, followed by integration
over $y$ and some simple manipulations, converts this expression
into
\begin{eqnarray}
\hspace*{-25mm} \phi_{\rm RS}
  &=&
  \frac{1}{2}-\frac{1}{4}
\frac{\int\!DxDz~\sgn[z\sqrt{q}+x\sqrt{1\minus q}+\beta
\sqrt{\alpha}(1\minus q)+\frac{m}{\sqrt{\alpha}}]
e^{\beta(z\sqrt{\alpha q}+ m) } \cosh^{n-1}[\beta(z\sqrt{\alpha
q}+m)]}{\int\!Dz~\cosh^{n}[\beta(z\sqrt{\alpha q}+m)]}
\hspace*{-20mm}\nonumber
\\
\hspace*{-25mm} && -\frac{1}{4}
\frac{\int\!DxDz~\sgn[z\sqrt{q}+x\sqrt{1\minus
q}+\beta\sqrt{\alpha}(1\minus q)-\frac{m}{\sqrt{\alpha}}]
e^{\beta(z\sqrt{\alpha q}- m)} \cosh^{n-1}[\beta(z\sqrt{\alpha
q}-m)]}{\int\!Dz~\cosh^{n}[\beta(z\sqrt{\alpha q}-m)]}
\hspace*{-20mm}\nonumber
\\
\hspace*{-25mm}
 &=&
  \frac{1}{2}-\frac{1}{4}
\frac{\int\!Dz~{\rm Erf}\left[\frac{z\sqrt{\alpha q }+\beta
\alpha(1\minus
q)+m}{\sqrt{2\alpha(1-q)}}\right]e^{\beta(z\sqrt{\alpha q}+ m) }
\cosh^{n-1}[\beta(z\sqrt{\alpha
q}+m)]}{\int\!Dz~\cosh^{n}[\beta(z\sqrt{\alpha q}+m)]} \nonumber
\\
\hspace*{-25mm} && -\frac{1}{4} \frac{\int\!Dz~{\rm Erf}
\left[\frac{z\sqrt{\alpha q}+\beta\alpha(1\minus
q)-m}{\sqrt{2\alpha(1-q)}}\right]e^{\beta(z\sqrt{\alpha q}- m) }
\cosh^{n-1}[\beta(z\sqrt{\alpha
q}-m)]}{\int\!Dz~\cosh^{n}[\beta(z\sqrt{\alpha q}-m)]}
\label{eq:phi_RS}
\end{eqnarray}
In the paramagnetic state, where $m=q=0$, this simplifies further
to
\begin{eqnarray}
\phi_{\rm RS}&=&
  \frac{1}{2}-\frac{1}{2}{\rm Erf}[\beta\sqrt{\frac{\alpha}{2}}]
  \label{eq:phi_P}
\end{eqnarray}
In the recall and spin-glass states the evaluation of
(\ref{eq:phi_RS}) requires the (numerical) solution of the RS
order parameters $\{m,q\}$ from  (\ref{eq:RSmsimple}). Examples of
the resulting curves as functions of temperature are shown in
figure \ref{fig:misaligned}, for $\alpha\in\{0.1,0.5,1.0\}$ and
$n\in\{1,2,3,4,5\}$. We note that for these values of $n$ replica
symmetry should be stable. The fraction of misaligned spins is
seen to decrease with increasing $n$ (i.e. with decreasing
connectivity temperature), as expected. This effect becomes more
pronounced for larger $\alpha$, where the amount of frustration to
be reduced by the geometry dynamics should indeed be largest.  In
the paramagnetic phase (large $T$) we see that $\phi_{\rm RS}$ is
independent of $n$, in accordance with (\ref{eq:phi_P}).

\subsection{Comparison with numerical simulations}

In order to perform numerical simulations, we need an explicit
stochastic dynamical equation for  updating of the network
geometry variables $\bc=\{c_{ij}\}$, which must approach the
appropriate Boltzmann equilibrium state characterized by the slow
Hamiltonian (\ref{eq:slowham}). Here we used a Glauber type Markov
process, where candidate bonds $c_{ij}$ are drawn randomly at each
iteration step  and then flipped with probability $W[F_{ij} \bc;
\bc]$, where $F_{ij}$ denotes the bond switch operator defined by
$F_{ij}c_{ij}=1-c_{ij}$, $F_{ij}c_{k\ell}=c_{k\ell}$ if $(i,j)\neq
(k,\ell)$:
\begin{equation}
W[F_{ij} \bc; \bc] =
\frac{1}{2}\left\{1-\tanh(\frac{\betaslow}{2}[ H_{\rm
s}(F_{ij}\bc)-H_{\rm s}(\bc)]) \right\}
\end{equation}
Detailed balance is built in. Upon inserting the slow Hamiltonian
(\ref{eq:slowham}) and using the scaling property
$\lim_{N\to\infty}c/N=0$  of our present extreme dilution regime,
one can for large $N$ rewrite our transition probabilities as
\begin{equation}
\hspace*{-15mm} W[F_{ij} \bc; \bc] = \frac{1}{2}\left\{ 1-
\tanh\left[ \frac{1}{2}(2c_{ij}-1)[\log\left(\frac{c}{N}\right) +
\frac{ \betaslow}{c} \sum_\mu \xi_i^\mu \xi_j^\mu \bra \sigma_i
\sigma_j \ket] \right] \right\} \label{eq:geometry_dynamics}
\end{equation}
where, as before, $\bra \ldots \ket$ indicates an equilibrium
average for the fast system, in Boltzmann equilibrium with
Hamiltonian (\ref{eq:fastham}), for a given  connectivity matrix
$\bc$.

In the present type of systems with multiple adiabatically
separated time-scales and nested equilibrations, the verification
of theoretical results by numerical simulations is known to be a
highly  demanding task. Even without the evolving geometry,  full
equilibration of the spins requires relaxation times which diverge
with $N$ faster than polynomially. If on top of this one aims to
also approach geometry equilibrium, which involves $\order(N^2)$
stochastic degrees of freedom, the system sizes accessible in
practice for numerical experimentation are small. Thus profound
finite size effects are unavoidable. It turns out that, when
simulating the process (\ref{eq:geometry_dynamics}), geometry
equilibration times are indeed extremely long, especially close to
phase transitions. This limits our ambitions regarding size, with
the standard CPU resources at our disposal, to the order of $N\sim
10^2$ spins.
 Since in our chosen scaling regime of extreme dilution we
have to simultaneously minimize $c^{-1}$ and $cN^{-1}$, we have in
our numerical experiments chosen $c=\sqrt{N}$.

\begin{figure}[t]
\vspace*{5mm} \hspace*{40mm}\setlength{\unitlength}{0.65mm}
\begin{picture}(200,100)
 \put(0,0){\epsfig{file=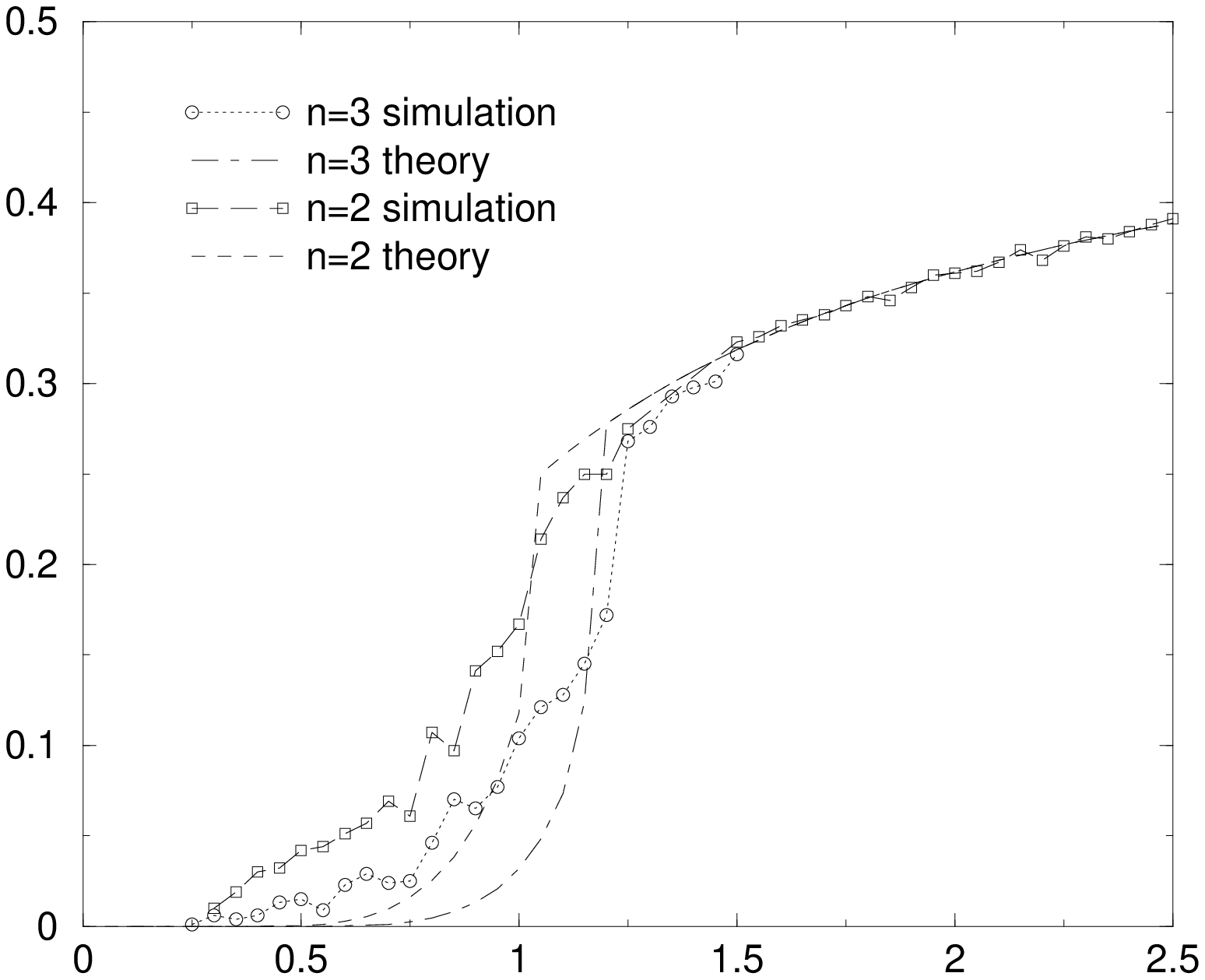, angle=0,
width=125\unitlength}} \put(-4,56){\small $\phi$}
\put(65,0){\small $T$}
\end{picture}
\vspace*{-3mm}
 \caption{Comparison between simulation measurements (all with $N=200$) and RS theoretical
predictions for the fraction of misaligned spins
$\phi=N^{-1}\sum_i\theta[-\sigma_i h_i]$ (where $h_i$ is the local
field at site $i$), as functions of temperature. The data shown
refer to  $\alpha=0.5$ with $n=2$ (simulations: connected squares;
theory: dashed lines) or $n=3$ (simulations: connected circles;
theory: dotted-dashed lines). Due to the need to equilibrate two
nested disordered processes, conventional computer  resources
limit experimentation to modest values of $N$. In spite of the
resulting finite size effects, the graph does  show satisfactory
qualitative agreement between theory and experiment. }
\label{fig:sims}
\end{figure}

Different macroscopic quantities could in principle be used for
testing our theory against experiments. The advantage of
observables such as $m$ and $\phi$ is that they can be measured
instantaneously, in contrast to the spin-glass order parameter
$q$. Here we have opted for the fraction of misaligned spins
$\phi$. The results are shown in \ref{fig:sims}, where we observe
qualitative agreement between theory and simulations. The
deviations observed in such experiments are found to decrease with
increasing system size $N$, albeit slowly.

\section{Stability of non-condensed retrieval states}

\label{sec:tausys}

The significant enlargement of the retrieval phase caused by our
geometry adaptation (see e.g. figures \ref{fig:phased} and
\ref{fig:intersect}) could have as a drawback that retrieval of
patterns other than the condensed one becomes impossible. Here we
address the question of whether the present `tayloring' of the
geometry variables $\{c_{ij}\}$ to one condensed state will leave
a finite basin of attraction for the non-condensed patterns, or
whether recalling the latter requires
 a rewiring of the system (e.g. by temporarily raising the temperature
$\tilde{T}$)  to undo the established geometry. For large $\alpha$
most retrieval states must be unstable for any given geometry in
the extreme dilution scaling regime, since optimal capacity
calculations a la Gardner for diluted networks predict a finite
storage capacity \cite{Shim}.

To answer our question we will  study a second (fast) spin system
of $N$ spins $\btau=\{\tau_i\}$, governed again by the fast
Hamiltonian (\ref{eq:fastham}), with patterns and geometry
identical to that of the first. In particular, the connectivity
statistics are again given by
\begin{equation}
P(\bc) = Z_{\rm s}^{-1} e^{-\betaslow H_{\rm s}(\bc)}
\label{distrc}
\end{equation}
The slow Hamiltonian $H_{\rm s}(\bc)$ continues to be defined in
terms of the original spins $\bsigma$, assumed in a condensed
state characterized by a finite overlap with the first pattern,
and will therefore be taylored towards the recall of that
particular pattern. By studying in the $\btau$ system the
properties of states which are condensed in patterns two or
higher, we gain access to the stability of non-condensed retrieval
states in the original $\bsigma$ system.

The geometry-averaged free energy per spin of our new system is
calculated by using the replica trick in its conventional form,
i.e. via
\begin{eqnarray}
[f_\tau]&=& -\lim_{\hat{n}\to 0}\lim_{N\to\infty}\frac{1}{\beta
\hat{n} N}\log\left\{ \sum_{\bc} P(\bc) \left[\sum_{\btau}
e^{-\beta H_{\rm f}(\btau,\bc)}\right]^{\hat{n}} \right\}
\nonumber
\\
&=& -\lim_{\hat{n}\to 0}\lim_{N\to\infty}\frac{1}{\beta \hat{n}
N}\log\left\{   Z_{\rm s}^{-1}
\sum_{\{\bsigma^\alpha\}}\sum_{\{\btau^\gamma\}}\sum_{\bc}
e^{-\log\left(\frac{N}{c}\right)\sum_{i<j} c_{ij}}
\right.\nonumber
\\
&&\left.\hspace*{35mm}\times~
  e^{\frac{\beta}{c}\sum_{i<j}c_{ij}(\bxi_i\cdot\bxi_j)[\sum_{\alpha=1}^{n} \sigma^\alpha_i \sigma^\alpha_j
+\sum_{\gamma=1}^{\hat{n}} \tau^\gamma_i \tau^\gamma_j]} \right\}
\end{eqnarray}
The  next stages of analysis are sufficiently similar to those
followed earlier to justify limiting ourselves to giving the final
result in RS approximation. If again we assume at most $r$
patterns to be condensed we find:
\begin{eqnarray}
 [f_\tau]^{\rm RS}&=& {\rm extr}_{\{\hat{m}_\mu,\hat{q},a\}}\left\{
 \frac{1}{2} \sum_{\mu\leq r} \hat{m}_\mu^2 -
\frac{1}{4}\alpha \beta (\hat{q}-1)^2 + \frac{1}{2}\alpha \beta n
a^2 - \frac{1}{\beta} \log 2 \right.\nonumber \\ && \left.
\hspace*{10mm} - \frac{1}{\beta} \left\bra \frac{\int\! Dy Dz~
\cosh^n(\Xi_1) \log \cosh (\Xi_2)}{\int\! Dy ~\cosh^n(\Xi_1)}
\right\ket_\bxi\right\}
\\
\Xi_1 &=& \beta\left(\bm \cdot \bxi + y\sqrt{\alpha q}\right) \\
\Xi_2 &=& \beta\left(\hat{\bm} \cdot \bxi +
\frac{a}{q}\sqrt{\alpha q} [y + z\sqrt{\hat{q}q/a^2 - 1}] \right)
\label{Guzais}
\end{eqnarray}
In addition to the previously encountered order parameters
$\{\bm,q\}$, which relate to the $\bsigma$ system (and continue to
be defined as the solution of the earlier saddle-point equations),
we now have new order parameters $\{\hat{\bm},\hat{q},a\}$, whose
physical meaning is found to be \bd \hat{m}_\mu =
\lim_{N\to\infty}\frac{1}{N}\sum_i\overline{\bra \xi^\mu_i
\tau_i\ket} ~~~~~~ \hat{q} =
\lim_{N\to\infty}\frac{1}{N}\sum_i\overline{\bra \tau_i\ket^2}
~~~~~~ a = \lim_{N\to\infty}\frac{1}{N}\sum_i\overline{\bra
\sigma_i \tau_i \ket} \ed The new order parameters are to be
solved from the saddle-point equations
\begin{eqnarray}
\hat{m}_\mu &=& \left\bra \xi_\mu \frac{\int\! Dy Dz~ \tanh(\Xi_2)
\cosh^n(\Xi_1)} {\int\! Dy ~\cosh^n(\Xi_1)} \right\ket_\bxi
\nonumber
\\ \hat{q} &=& \left\bra \frac{\int\! Dy Dz~ \tanh^2(\Xi_2)
\cosh^n(\Xi_1)} {\int\! Dy~ \cosh^n(\Xi_1)} \right\ket_\bxi
\nonumber
\\ a &=& \left\bra \frac{\int\! Dy Dz~ \tanh(\Xi_1)
\tanh(\Xi_2)\cosh^n(\Xi_1)} {\int\! Dy~ \cosh^n(\Xi_1)}
\right\ket_\bxi
\end{eqnarray}
It can be shown that solutions of these equations will obey $a^2
\leq \hat{q} q$ (to be expected in view of the square root in
$\Xi_2$).

\begin{figure}[t]
\vspace*{-5mm} \hspace*{5mm}
 \setlength{\unitlength}{0.58mm}
\begin{picture}(270,100)
 \put(-10,-10){\epsfig{file=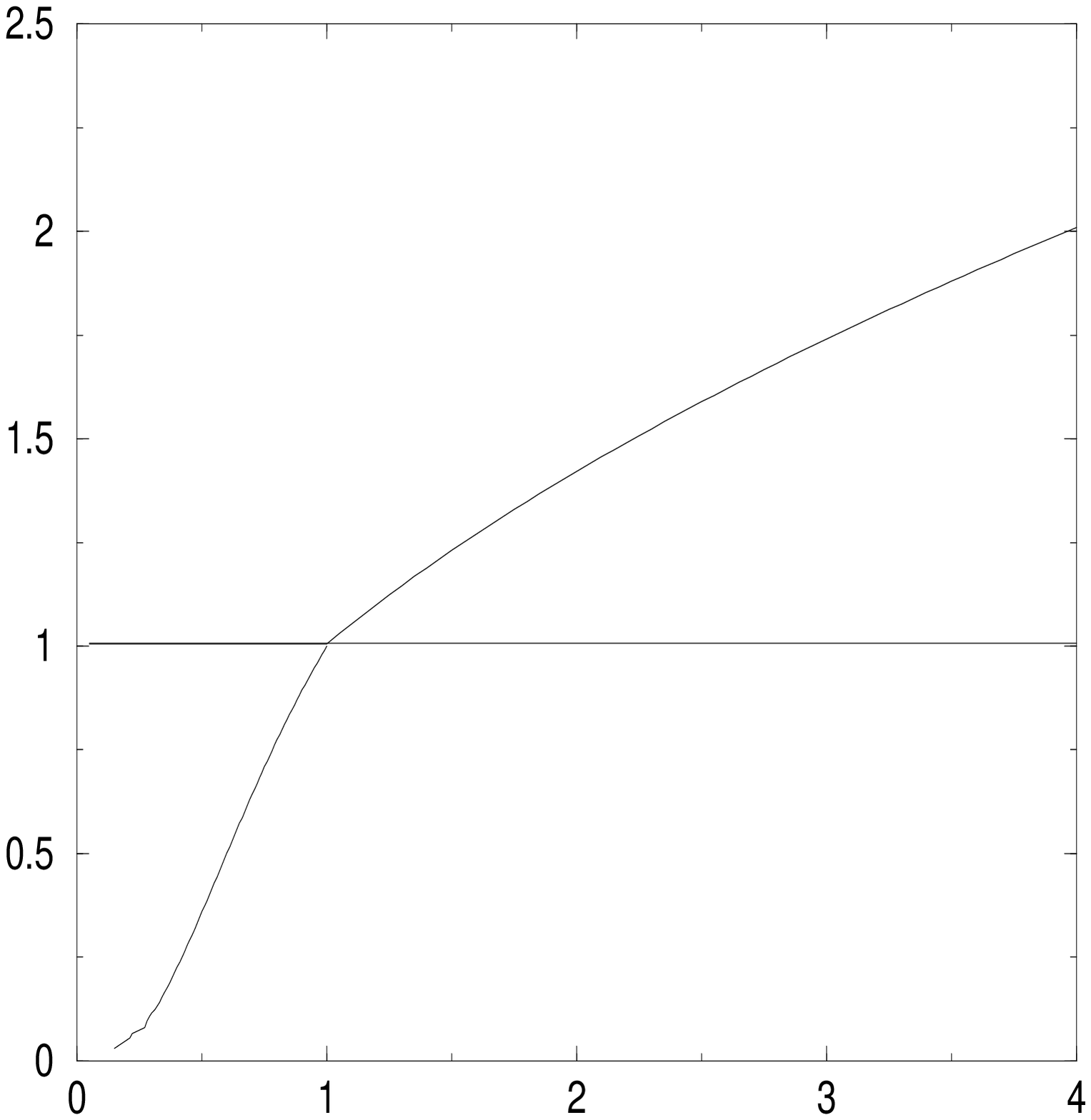, angle=0, width=90\unitlength}}
 \put(85,-10){\epsfig{file=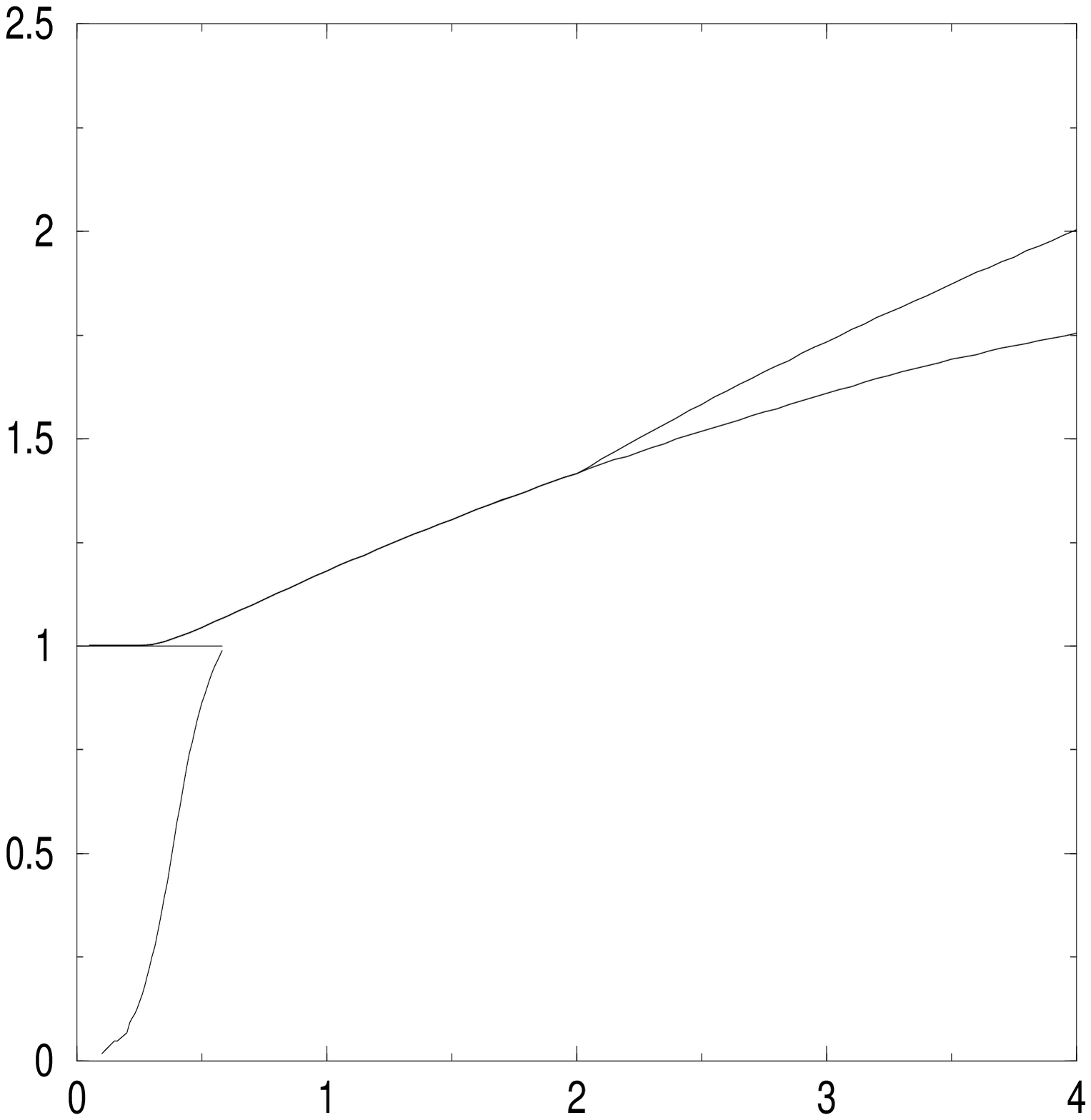, angle=0, width=90\unitlength}}
 \put(180,-10){\epsfig{file=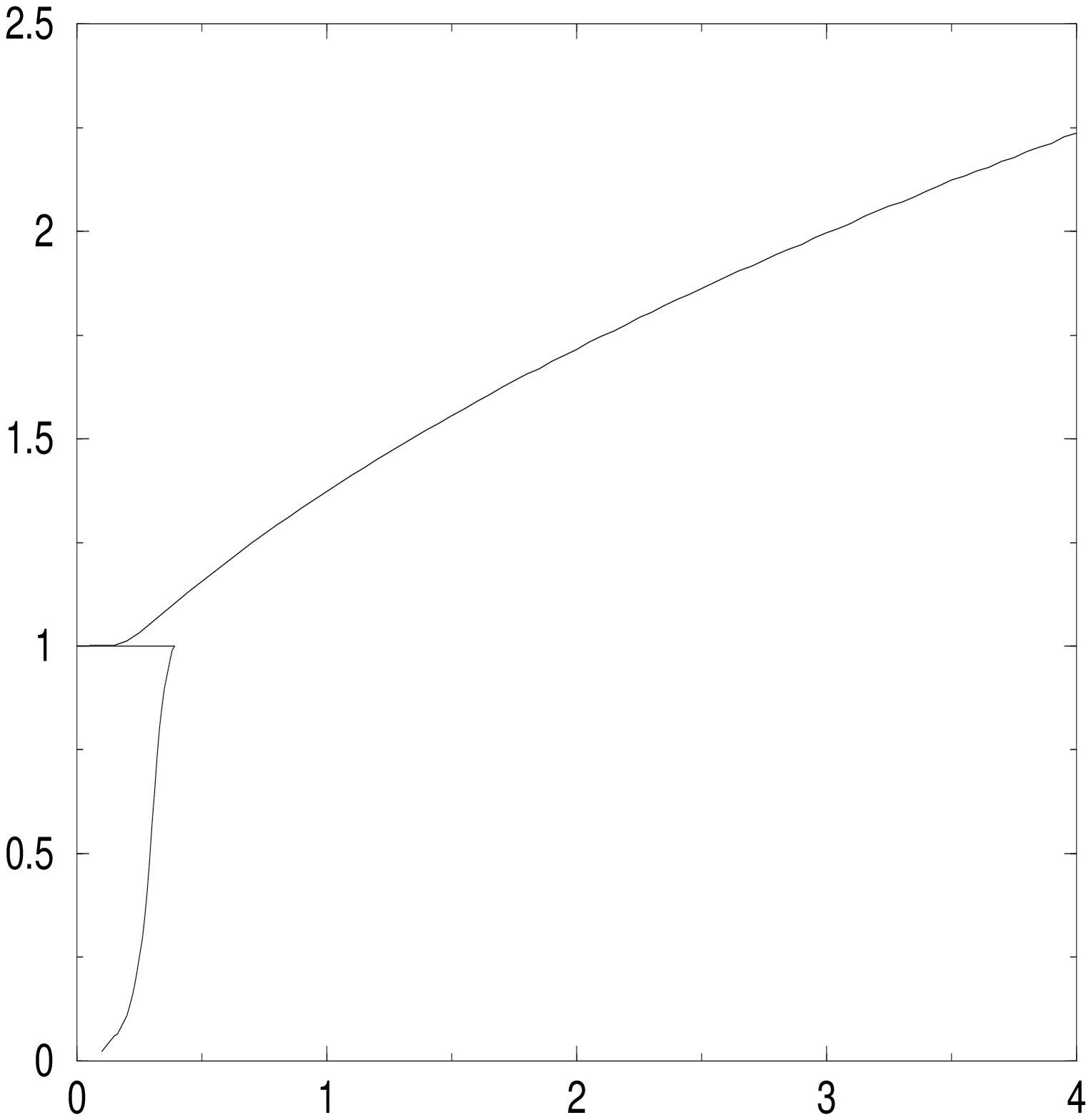, angle=0, width=90\unitlength}}
 \put(-20,39){\small $T$}
\put(37,-15){\small $\alpha$} \put(132,-15){\small $\alpha$}
\put(227,-15){\small $\alpha$} \put(50,77){\footnotesize $n=1$}
\put(145,77){\footnotesize $n=2$} \put(240,77){\footnotesize
$n=3$}

 \put(7,65){\footnotesize P}\put(102,65){\footnotesize
P}\put(197,65){\footnotesize P}
 \put(60,45){\footnotesize
SG}\put(165,59){\footnotesize SG}

 \put(50,15){\footnotesize R$_1$}\put(1,20){\footnotesize R$_2$}
 \put(145,15){\footnotesize R$_1$}\put(93,23){\footnotesize R$_2$}
  \put(240,15){\footnotesize R$_1$}\put(188,24){\footnotesize R$_2$}
\end{picture}
\vspace*{5mm} \caption{Cross-sections for fixed $n$ of the
expanded phase diagram, in which the previous retrieval phase R
has been separated into two sub-regions: ${\rm R}_1$ defines the
phase where only the nominated pattern can be recalled to which
the geometry has been taylored, and ${\rm R}_2$ defines the phase
where, in spite of the biased geometry,  all stored  patterns can
still be recovered. From left to right: $n=1,2,3$.}
\label{fig:tau}
\end{figure}

We now adopt a condensed ansatz which corresponds to the $\btau$
system being in a condensed state which differs from that of the
$\bsigma$ system (where the latter drives the geometry evolution):
$m_\mu=m\delta_{\mu 1}$,  $\hat{m}_\mu = \hat{m}\delta_{\mu 2}$.
Solutions of this type must have $a=0$, which is reasonable
considering that any finite correlation between the $\btau$ and
$\bsigma$ systems makes $\hat{m}_1=0$ highly improbable. For $a=0$
our two systems decouple, with the equations for $\hat{m}$ and
$\hat{q}$ reducing to
\begin{eqnarray}
\hat{m}&=& \int\! Dz~ \tanh(\beta[\hat{m} + z\sqrt{\alpha\hat{q}
}])
\\
\hat{q} &=& \int\! Dz~ \tanh^2(\beta[\hat{m} +
z\sqrt{\alpha\hat{q}}] )
\end{eqnarray}
These are exactly the RS equations of the model with frozen random
dilution \cite{WS}. The $\hat{m}_1=a=0$ solutions of our saddle
point equations could be unstable against perturbations in
$\hat{m}_1$ and $a$. In the paramagnetic phase, an expansion of
the free energy up to second order in the order parameters gives
\begin{eqnarray*}
 [f_\tau]^{\rm RS} ={\rm extr}_{\{\hat{\mu},\hat{q},a\}}\left\{ \frac{1}{2}(1-\beta)\sum_\mu \hat{m}_\mu^2 -
\frac{1}{4}\alpha \beta\hat{q}^2 + \frac{1}{2}\alpha \beta
a^2+~{\rm higher~orders}\right\}
\end{eqnarray*}
indicating that the physical solution of the saddle-point
equations is the one which minimizes the free energy with respect
to $\hat{\bm}$ and $a$, and maximizes it with respect to
$\hat{q}$, as is usual in the limit $\hat{n} \to 0$ \cite{MPV}.
Expansion around $\hat{m}_1=a=0$, with nonzero $\hat{m}_2$ and
$\hat{q}$, reveals that a second order instability in $a$ occurs
at the temperature
\begin{equation}
T_c = \sqrt{\alpha (1-\hat{q})[1+(n-1)q]} \label{eq:separator}
\end{equation}
Below $T_c$ the $\btau$ system will be captured in the
$\hat{m}_\mu =\hat{m} \delta_{\mu 1} $ state, with $\hat{m}_1 = m$
(so retrieval of states other than that to which the geometry has
adapted is impossible), whereas above $T_c$ the $\btau$ system can
be in a locally stable condensed state different from that in
which the $\bsigma$ system is found.  The free energy of the
$\hat{m}_\mu = m\delta_{\mu 1}$ state is, however, always lower
than that of other retrieval states, at any temperature. This
implies that in the latter states the $\btau$ system can be at
most locally stable. The line (\ref{eq:separator}) has been
calculated in RS ansatz; this seems reasonable,  since in the
model of \cite{WS} RSB does not occur for $n\geq 1$.  In figure
\ref{fig:tau} we show for $n\in\{1,2,3\}$ the line
(\ref{eq:separator}) which separates the previous retrieval phase
R into two sub-regions, one ${\rm R}_1$ where only recall of one
single pattern is possible (the one to which the geometry is
taylored), and a second region ${\rm R}_2$ where, in spite of the
biased geometry, multiple patterns can be recalled. As expected,
increasing $n$ (i.e. reducing the connectivity temperature, so the
`tayloring' of the geometry becomes more effective) reduces the
size of the ${\rm R}_2$ region.

\section{Conclusion}

We have studied extremely  diluted recurrent neural networks in
which the geometry is allowed to evolve on time-scales which are
adiabatically slower than the equilibration time of the (fast)
neurons. In contrast to earlier studies, the actual {\em values}
of the bonds remain frozen (they are here given by Hopfield's
\cite{Hopfield} recipe) and only the connectivity is dynamic,
which implies that the slow adaptation is reversible and will not
wipe out any stored information. Our motivation was to investigate
whether, by having a geometry dynamics which aims to reduce
frustration, the information retrieval properties of the system
can be improved. As in earlier models with slow bond dynamics
\cite{CoolenSherrington,PCS1,JBC,JABCP,CoolenUezu,FeldmanDotsenko,DotsFranzMezard}
the equilibrium properties of our model are described by a replica
theory with nonzero replica dimension $n$, where $n=
\betaslow/\beta$ is the ratio between the temperature of the
(fast) neurons and the temperature of the (slow) connectivity.

We have calculated phase diagrams, reflecting the stationary state
of the slowest stochastic system (i.e. the geometry). They reveal
a boosting of the retrieval phase, compared to the frozen
connectivity case, as soon as $n>0$. In fact, for nonzero $n$ the
storage capacity diverges at low temperatures, as long as $p \ll
N$. This at first sight somewhat surprising result is explained by
the observation that, in tayloring the geometry to the recall of a
single  condensed pattern, the system sacrifices the recall
quality of an infinite number of non-nominated patterns.
 RSB effects are as always
confined to small values of $n$ (below approx. 0.32).
 In order to measure the expected reduction in frustration as a result of the geometry dynamics
 we have calculated the
 fraction
of mis-aligned spins (where spin and local field are of opposite
sign). This fraction is indeed found to decrease with decreasing
temperature of the connectivity. In order to examine in which
region of the phase diagram retrieval states other than the
condensed pattern are still locally stable, we studied a pair of
identical diluted networks, both with the same Boltzmann type
connectivity distribution. The connectivity is taylored to reduce
frustration in only the first of the two copies. This allows one
to study scenarios corresponding to the recall of  patterns (in
the second copy) which are not the one to which the geometry is
adapted. Such recall is seen to be possible, but only in a
sub-region of the recall phase, whose size decreases with
increasing $n$.

It would appear an interesting question to examine to what extent
the properties of our model with slowly evolving geometry persist
in more (biologically) realistic scenarios, e.g. when the average
number of connections $c$ per neuron remains finite when
$N\to\infty$. Such studies will involve order-parameter functions,
see e.g. \cite{WemmenCoolen,IsaacNikos}, and require finite $n$
generalizations of finite connectivity replica theory.

\section *{Acknowledgment}

BW and NS acknowledge financial support from the FOM Foundation
(Fundamenteel Onderzoek der Materie, the Netherlands) and the
Ministerio de Educaci\'{o}n, Cultura y Deporte (Spain, grant
SB2002-0107).

\end{document}